%% file: main.tex
\documentclass[journal]{IEEEtran}
\input{macros} 
\begin{document}

\bstctlcite{IEEEexample:BSTcontrol}

\title{Joint Random Access and Localization in Cell-Free User-Centric Networks with Frequency-Selective Fading Channels}

\author{Simon~Tarboush,~\IEEEmembership{Student Member,~IEEE}, Eleni Gkiouzepi, and Giuseppe Caire,~\IEEEmembership{Fellow,~IEEE} 

\thanks{A preliminary version of this work has been accepted to the IEEE Global Communications Conference (GLOBECOM 2026)~\cite{Tarboush2026Random}. The authors are with the Communications and Information Theory Chair (CommIT), Faculty of Electrical Engineering and Computer Science, Technische Universit{\"a}t Berlin. Emails: {\{simon.tarboush,gkiouzepi,caire\}@tu-berlin.de}. The work of S. Tarboush was supported by the European Union, through the Horizon Europe Marie Skłodowska-Curie Doctoral Networks Programme “Intelligent sensing and communication as training network for perceptive mobile networks in 6G (ISAC-NEWTON)” under Grant 101169496. The work of E. Gkiouzepi and G. Caire was supported by the xG‑RIC project, part of the program Communication Systems “Souverän. Digital. Vernetzt.” (grant number 16KIS2429K) of the German Federal Ministry of Research, Technology, and Space (BMFTR).}
}

\markboth{This work has been submitted to the IEEE for possible publication. Copyright may be transferred without notice.}{draft}
\maketitle

\begin{abstract} 
We study random access (RACH) schemes for cell-free (CF) user-centric networks to handle many geographically distributed users with sporadic traffic and intermittent activity. The RACH must allow the system to: 1) detect preambles sent by the (yet unknown) random access users in the RACH slot; 2) localize them for fast allocation of user-centric radio-unit (RU) clusters. Most prior work uses simplified models, neglecting frame-synchronous but chip-asynchronous transmission, possible line-of-sight (LoS) propagation for certain user-RU pairs, and multipath non-line-of-sight (NLoS) propagation yielding frequency-selective channels. Building on our previous work, we consider location-dependent partitioned random access codebooks where users in a geographic area (location) use the corresponding subset of random access preambles. We present a unified framework for joint detection and localization over a spatially consistent network-wide channel model, incorporating these neglected aspects. We evaluate two schemes: 1) a ``legacy'' scheme using Zadoff-Chu (ZC) sequences, extending the 3GPP 2-step RACH to the CF case; 2) our multisource approximate message passing (AMP) approach extended to multipath frequency-selective fading. For both schemes, we develop novel approximated GLRT preamble detection and Maximum-Likelihood position estimators with super-resolution refinement, implicitly exploiting received signal strength, angle of arrival, and time-difference of arrival information embedded into LoS components. Numerical results show that the AMP-based scheme achieves superior preamble detection, while both schemes have similar and excellent localization capability.
\end{abstract}


\section{Introduction}
\label{sec:intro}

Next-generation wireless networks are expected to handle 
an unprecedented number of users with very diverse activity patterns. In particular, efficient random access low-latency schemes must be designed to handle massive populations of intermittently active devices with sporadic traffic \cite{You2021Towards,Liu2018Sparse}. 

Random access is an essential functionality in any wireless network, allowing idle and new users to join the network. However, currently deployed mechanisms exhibit fundamental limitations, such as large overhead and limited scalability~\cite{Liva2024Unsourced,Ozates2025Unsourced}. For example, the 3GPP two-step \ac{rach} mechanism~\cite{ETSI138} uses a set of \ac{zc} sequences~\cite{ETSI138,Liva2024Unsourced} (here referred to as {\em preambles}, or {\em codewords}) that users can transmit to signal their presence. At any given RACH slot, users select at random a preamble, send it, and append data in the corresponding push opportunity (PO), i.e., a block of the time-frequency resource following the RACH slot in the same frame. The random selection mitigates the problem of preamble collision, which would result into a collision (heavy interference) of the users' data in the same PO. A random access channel with a virtually unbounded number of users, where, at any time slot, only a finite number of users transmit making use of a common codebook, is referred to as \ac{ura}~\cite{Polyanskiy2017Perspective,Liva2024Unsourced,Ozates2025Unsourced} in the information theoretic literature. In fact, virtually any standardized \ac{rach} mechanism belongs to this class.

In conventional cellular architectures, a base station equipped with co-located massive \ac{mimo} arrays scans the \ac{rach} slots and detects the active preambles, which act as ``tokens'' signaling the presence of some user that aims to transmit data and possibly request an allocated transmission resource for subsequent connected traffic. Although \ac{ura} has been extensively investigated for massive MIMO systems with co-located antennas \cite{Chen2019Covariance,Liu2018Massive,Fengler2021NonBayesian,Fengler2022Pilot,Marata2025Activity,Tian2022Massive,Liu2024MLE}, extending the \ac{ura} principle to distributed \ac{cf} user-centric newtworks  \cite{Ngo2017Cell,Demir2021Foundations,Ngo2024Ultradense,Buzzi2026Why} is fundamentally more challenging because in such networks any user must be assigned a user-centric cluster of \acp{ru}. However, these clusters depend dynamically on the user's position, which is {\em a priori} unknown to the network for random access users. 
Moreover, the network doesn't know which subset of \acp{ru} to process to detect the preamble of any given active user until the user is detected, creating a circular dependency.

To solve this problem, the idea of a location-based partition of the random access codebook into disjoint subsets of codewords assigned to geographic regions (locations) of the network coverage area was originally proposed by our research group \cite{Cakmak2025Joint} and has since then become a popular research topic \cite{Okumus2025Type,Okumus2026Type,Krishnan2026Minimum}. 
This partition aims to resolve the problem of the unknown association between the codewords and the channel statistics (in particular, the \acp{lsfc} that depend on the distance from the unknown user position and the \acp{ru}). In the follow-up works \cite{Gkiouzepi2024Joint,Gkiouzepi2025Joint}, we showed that the active users' position determining the \acp{lsfc} can be accurately estimated by post-processing the output of the multisource \ac{amp} algorithm. The combination of location-based codebooks, \ac{amp} estimation and activity detection, and \acp{lsfc}-based active user positioning yields an efficient RACH scheme with fast user-centric cluster assignment for CF networks.  

A strong limitation of the current literature (e.g., \cite{Ganesan2021Clustering,Bai2022Activity,Jiang2023EMAMP, Zhang2024Activity,Zhang2025Harnessing,He2025RSS,Meng2025Network}, including our own work \cite{Cakmak2025Joint,Gkiouzepi2024Joint,Gkiouzepi2025Joint}) is that this has focused on simple, chip-synchronous, narrowband (i.e., frequency flat) channel models, where channels vectors are random zero-mean Gaussian vectors and spatial consistency is only given in terms of distance-dependent \acp{lsfc}. In this paper, we extend our work to a much more realistic propagation scenario, which fully accounts for frame-synchronous/chip-asynchronous transmission,\footnote{In modern 3GPP-like systems, random access takes place in a dedicated RACH slot in which contending users transmit their preamble (random access codeword), so frame-level synchronism is guaranteed by construction, while chip-level (sample-level) synchronism cannot be assumed because of imperfect timing advance and the unknown, user-dependent propagation delay.} frequency-selective multipath fading with both \ac{nlos} paths and, possibly, a \ac{los} path, where the latter contains the propagation delay from the random access user and a subset of the \acp{ru} in \ac{los}, as well as the \ac{aoa}. 

\subsection{Contributions}

To the best of the authors’ knowledge, this is the first work to develop a comprehensive, unified framework for uRA in CF networks under a realistic, spatially consistent, frequency-selective propagation environment, where a central processor jointly exploits observations collected across the \acp{ru} and performs activity detection and user localization. 
Spatial consistency implies that the user-RU channels naturally share a common scattering environment, leading to partially overlapping multipath structures. Furthermore, the proposed model fully accounts for the fact that, with respect to any given user location, \acp{ru} may have, in addition to the \ac{nlos} paths, a \ac{los} component. 
We consider two possible receiver pipelines. The first, closely aligned with the current ``legacy'' \ac{rach} design, operates in the time domain and exploits the favorable circular auto- and cross-correlation properties of \ac{zc} sequences, and it uses a \ac{glrt} approach to deal with the unknown delay of the LoS component in the activity detection. 
The second is an extension of the multisource \ac{amp} approach pioneered in~\cite{Cakmak2025Joint} to the asynchronous frequency-selective case and operates in the frequency domain using non-orthogonal codewords. The analytical characterization of the AMP output statistics is exploited to devise a \ac{glrt} detection and localization scheme.

For both RACH pipelines, the associated novel quasi-Maximum-Likelihood position estimator with super-resolution refinement can be interpreted as an (implicit) joint \ac{rss} (i.e., the \acp{lsfc}), \ac{aoa}, and \ac{tdoa} localization scheme. 

We validate our methods through extensive numerical simulations by comparing these two random access strategies under the same {\em channel load} (i.e., the average number of active random access users per unit area). The system performance is evaluated in terms of the \ac{fa} and \ac{md} probabilities of the random access users, together with the accuracy of position estimation.  
Numerical simulations show that the frequency-domain approach outperforms the time-domain approach in detection, while both schemes have a similar and excellent localization capability.

\subsection{Organization}

The remainder of this paper is organized as follows. Section \ref{sec:sys_ch_model} introduces the system and the proposed spatially consistent frequency-selective channel model. Section \ref{sec:rxsig_model} formulates the received signals for both time- and frequency-domain schemes. Section \ref{sec:td_pipeline} details the proposed receiver pipeline of the time-domain scheme, including preamble detection and active user localization. Section \ref{sec:fd_pipeline} details the proposed receiver pipeline of the frequency-domain scheme with analogous but technically different steps to achieve the same functionalities. Conclusions are pointed out in Section \ref{sec:conclusion}. 

{\bf Notation:} 
Beyond standard notation clear from the context, we use: $[N] \eqdef \{1, \ldots, N\}$ to indicate index sets; $\star$, $\circledast$, $\otimes$, and $\odot$ to indicate respectively continuous-time convolution, discrete-time cyclic convolution, Kronecker product, and Hadamard product; underlined boldfaced and boldfaced lower case letters to denote row and column vectors (e.g., $\underline{\av}$ and $\av$), and bold upper case letters to denote matrices (e.g., $\Am$), where we adopt the familiar Matlab-like notation $[\av]_n$, $[\Am]_{n,m}$, $[\Am]_{n,:}$ and $[\Am]_{:,m}$ to denote the $n$th element of a vector $\av$ and the $(n,m)$th element, the $\nth{n}$ row and the $\nth{m}$ column of a matrix $\Am$, respectively. Two-dimensional points on the real plane $\RR^2$ are denoted by sans-serif letters (e.g., $\xsf$).

\section{System and Channel Models}
\label{sec:sys_ch_model}

\subsection{System Model}
\begin{figure}[htb]
	\centering 	\includegraphics[width=0.35\textwidth]{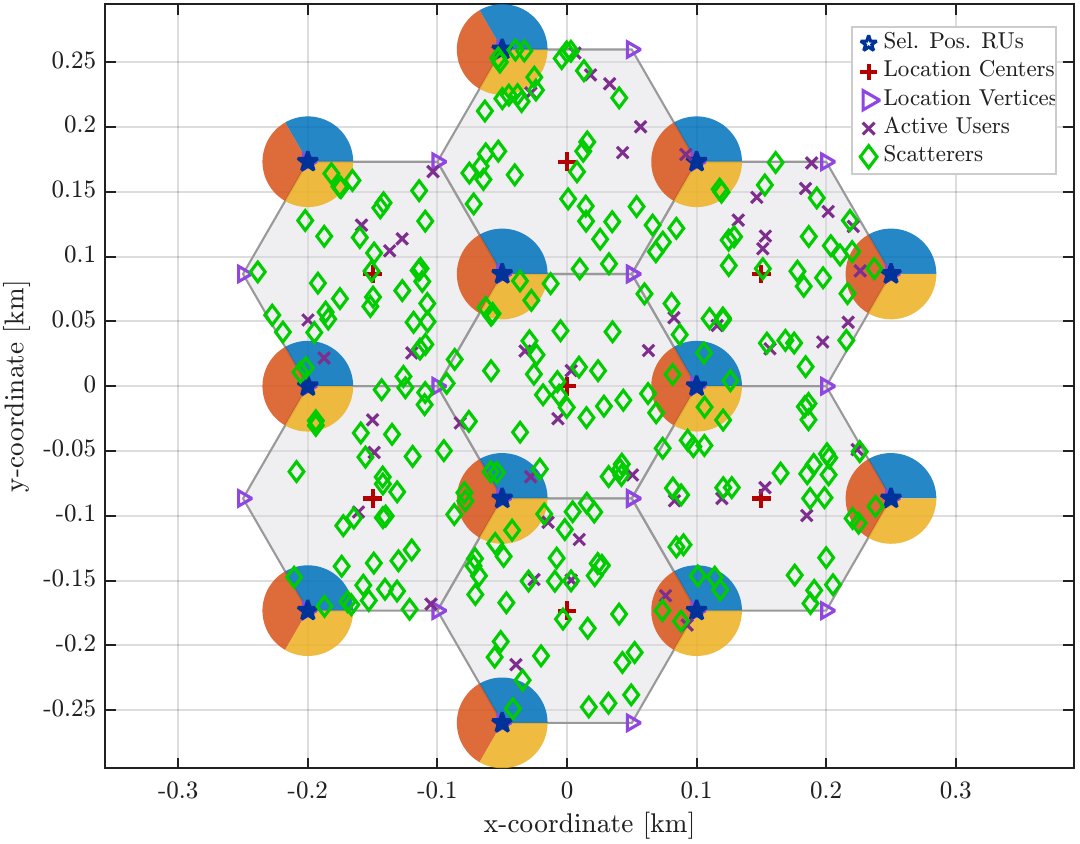}
	\caption{Example of the considered network showing the location of \acp{ru} and a random realization of 
    the active users and scatterers.}
\label{fig:network_topology}	
\end{figure}
We consider a \ac{cf} user-centric wireless network with $B$ \acp{ru} at fixed, known positions $\xsf_b \in \RR^2 : b \in [B]$, each equipped with $M$ elements \ac{ula}, serving a population of single-antenna random access users with unknown (to the network) positions $\ysf_k \in \Dc \subset \RR^2$, where $\Dc$ is the network coverage area.
In the same region $\Dc$, a total of $S$ scatterers are placed at positions $\zsf_s\in \RR^2 : s \in [S]$. 
For the purpose of the \ac{rach}, the coverage area $\Dc$ is partitioned into $U$ disjoint zones $\Dc_u: u \in [U]$, referred to as \textit{locations}.

While the system definition is general and the locations can be defined based on very accurate RU-dependent \acp{rm} \cite{Yapar2023Real,Jaensch2025Radio} or, more in general, \acp{ckm} \cite{Zeng2024Tutorial} (e.g., see \cite{Gkiouzepi2025Joint}), for the sake of numerical result reproducibility and ease of exposition in our numerical results we consider the classical hexagonal layout shown in Fig.~\ref{fig:network_topology}, where 12 sites are placed on a hexagonal lattice. Each site contains 3 \acp{ru}, each covering a 120$^\circ$ angle (indicated by the three colored sectors) for a total of $B = 36$ \acp{ru}. The coverage area (in gray) is partitioned into $U = 7$ locations $\Dc_u$ corresponding to the hexagons. Fig.~\ref{fig:network_topology} also shows a realization of the randomly placed active users and scatterers.

\subsection{Location-Based Unsourced Random Access}

The general \ac{rach} scheme works as follows \cite{Cakmak2025Joint,Gkiouzepi2024Joint,Gkiouzepi2025Joint}: each location $\Dc_u$ is assigned a distinct \ac{ura} codebook formed by $N_u$ codewords (\ac{rach} preambles) of length $L$ ``chips'' for a total duration of $\approx L T_\samp$ (without considering the use of \ac{cp}) where $T_\samp = 1/W$ and $W$ is the system bandwidth. The codewords are arranged for convenience as the columns of a codebook matrix $\Sm_u \in \CC^{L \times N_u}$, where $N$ is the total number of codewords in the system, $\sum_u N_u = N$.

Users can determine their location from trigger beacons transmitted periodically by all the \acp{ru} simultaneously to indicate the beginning of the \ac{rach} slot (e.g., by mapping the vector of received beacon powers from the different \acp{ru} to a location index as in \cite{Gkiouzepi2025Joint,Testi2026Weighted}). When a user in location $\Dc_u$ wishes to access the \ac{rach}, it chooses a codeword in codebook $\Sm_u$ and transmits it at a given power (fixed by the scheme) in the \ac{rach} slot. 
The system obtains the received signal in the \ac{rach} slot from all \acp{ru} and must identify the active codewords (i.e., the codewords that have been effectively transmitted) and the position of the corresponding active users with more accurate positioning with respect to the (typically very coarse) locations. The particular \ac{ura} codebook format and the corresponding detection and positioning algorithms constitute the main contribution of this paper.

The users may choose their codeword depending on some information bits \cite{Fengler2022Pilot} or simply at random \cite{ETSI138}. This aspect is irrelevant to our scope and shall not be discussed. Furthermore, some codewords may be chosen by more than one user. Such {\em preamble collisions} impact the overall \ac{rach} performance \cite{Liva2024Unsourced}, depending on the particular collision resolution scheme. Since this source of impairment is quite different in nature from the detection error events, in our results we assume that there is no collision. It is generally true that large codebooks can mitigate preamble collision and should be preferred. Schemes that estimate the codeword multiplicity, i.e., not only whether a codeword was or was not transmitted, but also how many users have transmitted the same codeword, have been studied under the denomination of ``type-based'' \ac{ura} \cite{Okumus2025Type,Okumus2026Type}.

\subsection{Spatially Consistent Channel Model}
\label{subsec:ch_model}

\subsubsection*{Continuous and discrete-time response} To illustrate the channel model, we indicate with the index $k$ a generic active user placed at $\ysf_k \in \Dc_u$, selecting a codeword $(u,n)$ ($n$th column of the matrix $\Sm_u$).\footnote{The $\nth{(u,n)}$ codeword is referred to as ``active'' if some random access user in location $\Dc_u$ transmits the $\nth{n}$ column $\sv_{u,n}$ of the associated codebook $\Sm_u$. In the following, we may refer to the user transmitting codeword $\sv_{u,n}$ as the {\em active user} $(u,n)$.}
The continuous-time single-input $M$-output channel (row vector) from user $k$ and \ac{ru} $b$ placed at $\pv_b$ is given by $\underline{\hv}_{b,k}(\tau) = \underline{\hv}^\los_{b,k}(\tau) + \underline{\hv}^\nlos_{b,k}(\tau)$. The \ac{los} component takes on the form
\begin{equation}
    \label{eq:los_comp}
    \underline{\hv}^\los_{b,k}(\tau) = \II^\los_{b,k} \sqrt{\beta^\los_{b,k}} {\rm e}^{j \phi^\los_{b,k}} \underline{\av}(\theta^\los_{b,k})g(\tau-\tau^\los_{b,k,0}),
\end{equation}
where $\beta^\los_{b,k}$ is the distance-dependent \ac{lsfc} of the \ac{los} path, $\phi^\los_{b,k}\!\sim\!\Uc[0,2\pi)$ is a uniform random phase to model the fact that any small change, even of the order of half a wavelength, in the user location may result in a significant phase-shift of the \ac{los} component~\cite{Ozdogan2019Performance,Chen2025Channel}, $g(\cdot)$ is the pulse shaping function, $\tau^\los_{b,k,0} = \|\xsf_b - \ysf_k\|/\csf$ is the propagation delay ($\csf$ is the speed of light), and $\underline{\av}(\theta^\los_{b,k}) \in \CC^{1 \times M}$ denotes the $\nth{b}$ \ac{ru} far-field array vector as a function of the local \ac{aoa} $\theta^\los_{b,k}$.\footnote{In this work we consider $\lambda/2$ spaced \acp{ula} at the \acp{ru}.} 
The local \ac{aoa} is defined with the boresight direction depending on the array orientation and the user and RU positions on the plane. 

The indicator function $\II^\los_{b,k} \in \{0,1\}$ in \eqref{eq:los_comp} controls the presence of the \ac{los} path. Specifically, the \ac{los} component from a user at $\ysf_k$ and a RU at $\xsf_b$ exists if the user is within a certain distance $d^{\rm LoS}_{\Rrm,\Urm}$ from the RU and the \ac{aoa} falls within the \ac{ru}'s angular sector $\Acal_b$). Formally $\II^\los_{b,k}$ is defined as
\begin{equation}
    \II^\los_{b,k}=\onebb\big( \|\xsf_b - \ysf_k\| \le d^{\rm LoS}_{\Rrm,\Urm}\;\land\; \theta_{b,k}\in\Acal_b\big).
\end{equation}
The \ac{nlos} component $\underline{\hv}^\nlos_{b,k}(\tau)$ is the sum of all single-bounce paths\footnote{Here we consider only single-bounce paths, as higher-order reflections typically, at relatively large communication distance, suffer significantly higher attenuation and contribute marginally to the received signal power.} generated by scatterers.
A \ac{nlos} path from a user at  $\ysf_k$ to a RU at $\xsf_b$ through a scatterer at $\zsf_s$ exists if 
\begin{equation}
\label{eq:pathexistence}
    \|\ysf_k - \zsf_s\| \le d^{\rm scat}_{\Srm,\Urm} \;\land\; \|\zsf_s - \xsf_b\| \le d^{\rm scat}_{\Rrm,\Srm} \;\land\; \theta_{b,s} \in \Acal_b,
\end{equation}
where $\theta_{b,s}$ is the \ac{aoa} of scatterer $s$ with respect to RU $b$.  
The truncation distances $d^{\rm LoS}_{\Rrm,\Urm}$, $d^{\rm scat}_{\Srm,\Urm}$, and $d^{\rm scat}_{\Rrm,\Srm}$ are given system parameters. Let $\Pc_{b,k} \subseteq [S]$ denote the set of NLoS paths satisfying \eqref{eq:pathexistence}. Then, the \ac{nlos} component between user $k$ and \ac{ru} $b$ is given by 
\begin{equation}
\label{eq:nlos_comp}
    \underline{\hv}^\nlos_{b,k}(\tau) =\sum_{p \in \Pc_{b,k}} \sqrt{\beta_{b,k,p}^\nlos} \rho_{b,k,p} \underline{\av}(\theta^\nlos_{b,p}) g(\tau-\tau^\nlos_{b,k,p}),
\end{equation}
where $\beta_{b,k,p}^\nlos$ is the pathloss along the $p$th path accounting for distance and scattering cross-section, $\rho_{b,k,p} \sim_{\iid} \Cc\Nc(0,1)$ (mutually independent for different indices), consistent with the Swerling-II scattering model \cite{Fishler2006Spatial}, and $\tau^\nlos_{b,k,p}$ is the path delay. Notice that, for both \ac{los} and \ac{nlos} components, delays, \acp{lsfc}, and \ac{aoa} are functions of the user's position (and scatterers) w.r.t. the \ac{ru}. Consequently, the adopted channel modeling ensures spatial consistency throughout the network. 

The pulse shaping $g(\tau)$ is a Nyquist pulse obtained by transmitting a given chip pulse and applying a chip \ac{mf} at the receiver. In particular, here we assume a rectangular unit-energy chip pulse $\psi(\tau)= \sqrt{W}\mathrm{rect}(\tau W-\frac{1}{2})$ with support $[0,1/W]$, such that the convolution $g = \psi \star \psi$ is a triangular pulse of support $[0,2/W]$.\footnote{All the results and approaches in this paper can be generalized to other Nyquist pulses. However, the mathematics becomes more cumbersome. In addition, for large bandwidth $W$, any reasonable Nyquist pulse can be well-approximated by a triangular pulse for the sake of the receiver processing.} 
The discrete-time channel $\underline{\hv}[\ell]$ is obtained by sampling $\underline{\hv}(\tau)$ at the chip rate $W = 1/T_s$. 
The triangular pulse delayed by $\tau_p$ and sampled at integer multiples of $T_s$ yields the discrete-time signal 
\begin{equation}
\label{eq:g_sampled}
  g\!\left(\!\frac{\ell - \tau_p W}{W}\!\right)
  \!=\!
  \begin{cases}
    \ell - \tau_p W, & \ell \in [\tau_p W,\;\tau_p W+1), \\
    2 + \tau_p W - \ell,& \ell \in [\tau_p W+1,\;\tau_p W+2], \\
    0, & \text{otherwise,}
  \end{cases}
\end{equation}
such that each path contributes at most two non-zero chip-spaced taps. 
Using \eqref{eq:g_sampled}, for a generic propagation path $p$, the integer delay is $\ell_{p} = \lceil \tau_{p} W \rceil$ and the fractional delay is $\mu_{p} = \ell_{p} - \tau_{p} W$, such that $\mu_{p} \in [0, 1)$. In particular, without loss of generality, by ordering delays as $\tau_0^\los < \tau_1^\nlos < \cdots < \tau_P^\nlos$, the minimum integer delay is given by  $\ell_0 = \lceil \tau_0^\los W \rceil$ and the maximum integer delay is $\ell_{\max} = \lceil \tau_P^\nlos W \rceil + 1$. As a consequence, the discrete-time channel length is $D+1$ with $D = \ell_{\max} - \ell_0$ and the channel has non-zero taps only for $\ell \in \{\ell_0,\ldots,\ell_0+D\}$.

\subsubsection*{Time-space channel matrix}
We define the $L \times 1$ vector of the shifted Nyquist pulse samples $\gv_{\mu_p}$, with elements as in \eqref{eq:g_sampled}, shifted so that the first non-zero coefficient is at the top and zero-padded to length $L$
\begin{equation}
    \label{eq:delay_int_frac}
    \gv_{\mu_p} = [\mu_p, 1-\mu_p, 0,\ldots,0]^\transp,
\end{equation}
where $\gv_{\mu_p}$ has at most two non-zero taps at its top.
The $L\times M$ time-space channel matrix, after zero padding to length $L$, using \eqref{eq:los_comp}, \eqref{eq:nlos_comp}, \eqref{eq:delay_int_frac}, can be written as
\begin{equation}
    \label{eq:discrete_time_space_ch_mat}
    \Hm^\circ \!=\! \II^\los \!\sqrt{\beta^\los} {\rm e}^{j \phi^\los} \Jm^{\ell_0} \gv_{\mu_0} \underline{\av}(\theta^\los)\! +\! \sum_{p \in \Pc} \sqrt{\beta_p} \rho_p \Jm^{\ell_p} \gv_{\mu_p}
\underline{\av}(\theta_p).
\end{equation}
where $\Jm$ is the $L\times L$ cyclic downshift matrix (see Appendix-\ref{app:circ_dft} for the definition of $\Jm$ \eqref{eq:cyc_shift} and the properties used). Each term of the type $\Jm^{\ell_p} \gv_{\mu_p} \underline{\av}(\theta_p)$ is a rank-one $L \times M$ matrix with at most only two non-zero rows in positions $\ell_p$ and $\ell_p + 1$. Moreover, we have $\ell_p \ge \ell_0$ for every $p \in \Pc$ since the LoS path is shorter than any NLoS path. 
Hence $\Jm^{\ell_p} = \Jm^{\ell_0}\Jm^{\ell_p-\ell_0}$ for every term in \eqref{eq:discrete_time_space_ch_mat}, and $\Jm^{\ell_0}$ factors out as
\begin{equation}
    \label{eq:ch_mat_factored}
    \Hm^\circ = \Jm^{\ell_0} \Hm,
\end{equation}
\begin{equation}
    \label{eq:ch_mat_aligned}
    \Hm = \II^\los \sqrt{\beta^\los} {\rm e}^{j\phi^\los} \gv_{\mu_0}\underline{\av}(\theta^\los) + \sum_{p\in\Pc}\sqrt{\beta_p}\rho_p\, \Jm^{\ell_p-\ell_0}\gv_{\mu_p}\underline{\av}(\theta_p)
\end{equation}
is the channel matrix re-referenced to delay $\ell_0$, with support collected in the top $D+1$ rows. The general form is
\begin{equation}
    \label{eq:alltop_matrix}
    \Hm = \begin{bmatrix} \underline{\hv}[\ell_0] \\ \underline{\hv}[\ell_0+1] \\ \vdots \\ \underline{\hv}[\ell_0+D] \\ \zerov \\ \vdots \\ \zerov \end{bmatrix},
\end{equation}
where each row $\underline{\hv}[\ell]$ contains the channel $\nth{\ell}$ tap coefficients of the $M$ RU antennas.\footnote{Row $0$ is guaranteed non-zero only when $\II^\los=1$; if the \ac{los} path is blocked, row $0$ and possibly a few rows after it may be identically zero until the first surviving \ac{nlos} tap.} In this form $\Hm$ carries only the small-scale fading and \acp{aoa} of the link, while the link's absolute range information is carried entirely by the single deterministic integer $\ell_0$ through $\Jm^{\ell_0}$ in \eqref{eq:ch_mat_factored}. Finally, the only difference between the \ac{los} and \ac{nlos} case is the statistics of the shortest-path coefficient, which in the \ac{los} case is non-Gaussian deterministic amplitude with random phase $\sqrt{\beta^\los} {\rm e}^{j\phi^\los}$ and in the \ac{nlos} case $(\sqrt{\beta_p}\rho_p)$ is zero-mean complex Gaussian. 

The time- and frequency-domain schemes for RACH detection and user positioning use different sequence lengths, $L \! = \! L_t \!\geq \!D+\!1$ and $L \! =\! L_f \!\geq \!D+\!1$, respectively. For notational simplicity, we use the same symbol $\Hm$ to indicate matrices of the form \eqref{eq:alltop_matrix} where the top $D+\!1$ rows are fixed by the channel and the remaining $L\! -\! D \!- \!1$ rows are zero-padded, where the cases $L \!= \!L_t$ or $L \!= \!L_f$ are clear from the context. 

\section{Time- and Frequency-Domain Received Signals}
\label{sec:rxsig_model}

\subsection{Time-domain ZC Signaling with CP}

In this section we develop the signal model for the considered time-domain scheme. 
We omit the user and the \ac{ru} antenna indices for notational simplicity. 
As anticipated before, the preambles are length-$L_t$ sequences $\sv$ from a \ac{zc} family with $L_t \gg D$. 
To ensure circular convolution in the presence of the multipath channel, a CP of length $L_\cp^\td$ with $L_\cp^\td \geq D $ is used, with total transmission duration $(L_t + L_\cp^\td)/W$ seconds.
The receiver discards the first $L_\cp^\td$ samples and obtains a block of length-$L_t$ affected by circular convolution with the time-domain channel impulse responses for each RU antenna. 
Stacking the received signal of all $M$ antennas by columns, the resulting  time-space received signal for the \ac{simo} channel from one user to one \ac{ru} with $M$ antennas is given by 
\begin{equation}
    \Ym = \Circ(\sv) \Jm^{\ell_0} \Hm.
    \label{eq:y_td_mimo}
\end{equation}
where $\Circ(\sv)$ is the circulant matrix build on $\sv$ (see Appendix-\ref{app:circ_dft}, definition \eqref{eq:cir_mat} and property \eqref{eq:conv_delayedJ}). 

\subsection{Time-domain RACH Received Signal}
\label{subsec:td_cf_rach_rxsig}

Using \eqref{eq:y_td_mimo}, the $L_t \times M$ time-domain signal matrix received at \ac{ru} $b$ in the \ac{rach} slot is given by 
\begin{equation} 
    \Ym_b = \sum_{u=1}^U \sum_{n=1}^{N_u} a_{u,n} \Circ(\sv_{u,n}) \Jm^{\ell_{b,u,n}} \Hm_{b,u,n}  + \Wm_b
    \label{eq:y_td_cell_free}
\end{equation}
where $a_{u,n}\in\{0,1\}$ is a Bernoulli-$\lambda_u$ activity variable with probability $\lambda_u \in (0,1)$ indicating whether codeword $(u,n)$ has been transmitted, the delay of the shortest path between the position of the user $(u,n)$ and the $\nth{b}$ \ac{ru} and the corresponding time-space matrix in the form \eqref{eq:alltop_matrix} are denoted by $\ell_{b,u,n}$ and $\Hm_{b,u,n}$, respectively, and $\Wm_b \in \CC^{L_t \times M} \sim_{i.i.d} \Cc\Nc(0, N_0)$ is the \ac{awgn} matrix with per-component variance $N_0$.\footnote{It is known that when the chip MF is a unit-energy square-root Nyquist pulse, the chip-rate sampled Gaussian noise is i.i.d. with components $\!\sim\! \Cc\Nc(0,N_0)$, where $N_0$ is the complex baseband noise power spectral density.} Notice that $\ell_{b,u,n}$ and $\Hm_{b,u,n}$ depend on the positions of active user $(u,n)$ and \ac{ru} $b$ via the channel model defined in Section \ref{subsec:ch_model}. In particular, the geometric information of the \ac{los} (if present) is contained in the top two rows of $\Hm_{b,u,n}$. Moreover, for all $(u,n)$ for which $a_{u,n} = 0$ (inactive codewords), the corresponding channel parameters are irrelevant, since their contribution to the received signal is zero. 

\subsection{Frequency-domain CP-OFDM Signaling}

For the frequency-domain scheme, we adopt \ac{cp}-\ac{ofdm} with $L_f$ subcarriers and subcarrier spacing $W/L_f$. The subcarrier index is $\xi= 0, \ldots, L_f-1$. An \ac{ofdm} symbol consists of $L_f$ frequency-domain symbols, corresponding to $L_f$ time-domain ``chips'' plus a CP of length $L_\cp^\fd \geq D$. The duration of an \ac{ofdm} symbol is therefore $(L_f + L_\cp^\fd)/W$ seconds.
The multisource \ac{amp} approach~\cite{Cakmak2025Joint} (details in Section \ref{subsec:fd-denoiser}), requires the transmission of a sequence of $Q$ \ac{cp}-\ac{ofdm} symbols. 
Using a jargon familiar in the radar literature, the OFDM symbol index $q$ is referred to as {\em slow-time} and the $L_f$ chips inside each OFDM symbols ticks at the {\em fast-time}.  The overall RACH preamble transmission length in chips is given by $Q(L_f  + L_\cp^\fd)$. For a fair comparison between the time-domain and the frequency-domain schemes, we impose $L_t + L_\cp^\td \approx Q(L_f + L_\cp^\fd)$. Since the multisource \ac{amp} algorithm works well for large block length $Q$, we shall use $L_f$ as small as possible, i.e., $L_f = D+1$ with CP length $L_\cp^\fd = D$. Using the minimal CP length $L_\cp^\td = D$ also for the time-domain approach, we find $L_t \approx Q(2D+1) - D$, where the $\approx$ is due to the fact that $L_t$ is constrained by the ZC construction while $Q$ is arbitrary. 

The received signal corresponding to the transmission of one OFDM symbol in the time domain after CP elimination takes on the same form as \eqref{eq:y_td_mimo} with length $L_t$ replaced by length $L_f$.
The frequency-domain received signal is obtained by applying a unitary $L_f \times L_f$ \ac{dft} matrix $\Fm$ to \eqref{eq:y_td_mimo}, yielding 
\begin{equation}
    \tilde{\Ym} = \diag(\tilde{\sv}) \Phim^{\ell_0} \check{\Hm},
    \label{eq:y_fd_mimo_norm}
\end{equation}
where $\check{\Hm}\! =\! \sqrt{L_f} \Fm \Hm$ is the standard $L_f$-points DFT of the columns of $\Hm$, $\tilde{\sv} \!=\! \Fm \sv$ is the $L_f$-subcarriers frequency-domain OFDM symbol corresponding to the length-$L_f$ time-domain block $\sv$, and $\Phim$ is the linear-phase shift matrix (see Appendix-\ref{app:circ_dft} for the definitions of $\Fm$, $\Phim$, \eqref{eq:cir_mat_dft}, and property \eqref{prop:F_Phi_J}).

\subsection{Frequency-domain CF RACH Received Signal}
\label{subsec:fd_cf_rach_rxsig}

Using \eqref{eq:y_fd_mimo_norm} for the $\nth{q}$ \ac{ofdm} symbol $(q \in [Q])$, the received $L_f \times M$ frequency-space signal matrix at \ac{ru} $b$ is 
\begin{equation}
    \tilde{\Ym}^{(q)}_b = \sum_{u=1}^U \sum_{n=1}^{N_u} a_{u,n} \diag(\tilde{\sv}_{u,n}^{(q)}) \Phim^{\ell_{b,u,n}} \check{\Hm}_{b,u,n}  + \tilde{\Wm}^{(q)}_b,
    \label{eq:y_fd_cell_free_qOFDM}
\end{equation}
where $\tilde{\sv}_{u,n}^{(q)}$ is the $q$th frequency-domain OFDM symbol transmitted by user $(u,n)$, $\check{\Hm}_{b,u,n}$ is the corresponding frequency-space channel matrix, and the frequency-domain \ac{awgn} $\tilde{\Wm}^{(q)}_b$ is $\sim_{i.i.d} \Cc\Nc(0, N_0)$ since it is related by a unitary transform to the time-domain \ac{awgn}.
The $Q$ OFDM symbols $\{\tilde{\Ym}^{(q)}_b \!:\! q \in\![Q]\}$ can be arranged in a $L_f\! \times\! M \times\! Q$ subcarriers$\times$antennas$\times$slow-time ``data cube''. For each subcarrier $\xi\!=\!0,\!\ldots\!,\!L_f\!-\!1$, the corresponding $Q\!\times \!M$ slow-time$\times$antennas slice yields the received signal in the form
\begin{equation}
    \tilde{\Ym}_b[\xi] = \sum_{u=1}^U \sum_{n=1}^{N_u} a_{u,n}   \tilde{\sv}_{u,n}[\xi] 
    \breve{\underline{\hv}}_{b,u,n}[\xi]  + \tilde{\Wm}_b[\xi],
\label{eq:y_fd_cell_free_xi_subc}
\end{equation}
where $\tilde{\sv}_{u,n}[\xi] \in \CC^Q$ is the (column) vector of the frequency-domain \ac{ura} codeword transmitted on subcarrier $\xi$, and we define the frequency-domain channel vectors
\begin{equation}
\underline{\breve\hv}_{b,u,n}[\xi] \eqdef {\rm e}^{-j\frac{2\pi \xi}{L_f}\ell_{b,u,n}} [\check\Hm_{b,u,n}]_{\xi,:};
\label{hbreve}
\end{equation}
Here, for the sake of AMP activity detection, the delay-induced phase term ${\rm e}^{-j\frac{2\pi}{L_f}\ell_{b,u,n}\xi}$ is absorbed into the channel $\underline{\breve\hv}_{b,u,n}[\xi]$. However, this information will be exploited for positioning in Section \ref{subsec:fd-glrt-pos}. Subsequently, we define $\Xm_{b,u}[\xi] \in \CC^{N_u \times M}$ with rows $\big[\Xm_{b,u}[\xi]\big]_{n,:} = a_{u,n} \breve{\underline{\hv}}_{b,u,n}[\xi]$ for $n \in [N_u]$. Stacking the $N_u\times M$ blocks $\Xm_{b,u}[\xi]$ across all $B$ \acp{ru} into $\Xm_u[\xi]=(\Xm_{1,u}[\xi],\ldots,\Xm_{B,u}[\xi])\in\CC^{N_u\times F}$, $F= MB$, yields the per-subcarrier observation model
\begin{equation}
    \tilde\Ym[\xi] = \sum_{u=1}^U\tilde\Sm_u[\xi]\,\Xm_u[\xi] + \tilde\Wm[\xi],
\label{eq:y_fd_cell_free_total_subc}
\end{equation}
with $\tilde\Ym[\xi]= \big[\tilde{\Ym}_1[\xi], \tilde{\Ym}_2[\xi], \ldots, \tilde{\Ym}_B[\xi]\big]\in\CC^{Q\times F}$, $\tilde{\Wm}[\xi]= \big[\tilde{\Wm}_1[\xi], \tilde{\Wm}_2[\xi], \ldots, \tilde{\Wm}_B[\xi] \big]$, and $\tilde\Sm_u[\xi]\in\CC^{Q\times N_u}$ collecting the columns $\tilde\sv_{u,n}[\xi]$.

\section{Time-Domain Scheme for Joint Activity Detection and User Positioning}
\label{sec:td_pipeline}

The time-domain receiver consists of a bank of \acp{mf} at each RU, followed by a \ac{glrt} \cite{Poor2013Introduction} activity detection and novel localization algorithm restricted to the users detected as {\em active}. For each location $\Dc_u$, the \ac{glrt} activity detection and localization make use of the MF outputs at subsets of RUs conveniently located with respect to $\Dc_u$.

\subsection{Matched Filter Receiver}
\label{subsec:td-mf}

The cyclic cross-correlation of two codewords $\sv_{u,n}$ and $\sv_{v,z}$ is given by $\rv_{u,n,v,z} = \Circ(\sv_{u,n})^\herm \sv_{v,z}$. For $(u,n) = (v,z)$ this yields the cyclic autocorrelation of $\sv_{u,n}$. 
\ac{zc} sequences have quasi-ideal cyclic autocorrelation and cross-correlation properties, namely,  
\begin{equation}
\begin{aligned}
    \rv_{u,n,u,n}&\approx \Ec_{\rm zc}\,\ev_0, \\ \rv_{u,n,v,z}&\approx\zerov \;\;\forall (u,n)\ne(v,z),
\end{aligned}
\label{eq:ideal_corr}
\end{equation}
where $\Ec_{\rm zc} = \|\sv_{u,n}\|^2 = L_t \Ec_s$ is the total energy of the preamble sequence ($\Ec_s = P_{\mathrm{tx}}/W$ is the per-chip energy corresponding to transmit power $P_{\rm tx}$ in Watts), $\ev_\ell$ the $\nth{\ell}$ 1-hot standard basis versor. 
The (discrete-time) unit-energy \ac{mf} filter for $\sv_{u,n}$ ar RU $b$ corresponds to  left-multiply the received time-space signal $\Ym_b$ in \eqref{eq:y_td_cell_free} by $\frac{1}{\sqrt{\Ec_{\rm zc}}} \Circ(\sv_{u,n})^\herm$. 
Using \eqref{eq:ideal_corr} with $\Circ(\ev_0)=\Id_{L_t}$, leads to
\begin{equation}
    \Ym^{\mf}_{b,u,n} \approx a_{u,n}\sqrt{\Ec_{\rm zc}}\,\Jm^{\ell_{b,u,n}}\Hm_{b,u,n} + \Wm''_{b,u,n},
\label{eq:mf_out_td}
\end{equation}
where $\Wm''_{b,u,n}$ collects the Gaussian noise and the residual multi-user interference due to not exactly zero cross-correlations in \eqref{eq:ideal_corr}, where $\EE[\Wm''_{b,u,n}(\Wm''_{b,u,n})^\herm]\succeq N_0\Id_{L_t}$ (in the positive semi-definite order). We write its per-entry variance as $\sigma_w \ge N_0$, which is a nuisance parameter that absorbs residual interference and is learnable offline (details in Section \ref{subsec:radio_map}).

\subsection{Activity Detection and Coarse Positioning using GLRT}
\label{subsec:td-glrt}

Thanks to the location-based partitioning of the \ac{ura} codebook, a centralized processor knows that if codeword $(u,n)$ is active, it is received at a strong signal level at a subset $\Vc_u \subseteq [B]$ of \acp{ru} conveniently located with respect to location $\Dc_u$. In particular, $\Vc_u$ contains the \acp{ru} with \ac{los} path to location $\Dc_u$ plus possibly some other RU. This association is used to control complexity and to improve the mismatched-likelihood approximation by discarding \acp{ru} that see only weak \ac{nlos} signal energy.

Conditioned on activity $(a_{u,n}=1)$ and on a postulated $(u,n)$ active user position $\ysf \in \Dc_u$, the unknown delays $\{\ell_{b,u,n}:b\in\Vc_u\}$ are all determined via the purely geometric function $\ell_{b,u,n}(\ysf) =\lceil|\ysf-\xsf_b|W/\csf\rceil$. 
Hence, the observation can be re-parameterized and expressed directly in terms of $\ysf$ ranging over a discretization grid $\Gc_u$ of $\Dc_u$.\footnote{The grid design is discussed in Section \ref{subsec:grid_pos}.} 
Activity detection based on the joint observations $\{\Ym^{\mf}_{b,u,n} : b \in \Vc_b \}$ is a {\em composite binary hypothesis test} \cite{Poor2013Introduction}, where the null hypothesis $\Hc_{0}$ corresponds to a fixed distribution of the observation (in our case, the distribution of $\Ym^{\mf}_{b,u,n}$ in \eqref{eq:mf_out_td} for $a_{u,n} = 0$, containing only noise), and the alternative hypothesis $\Hc_{1}$ corresponds to a family of distributions for all the values of some parameter (in our case, the distribution of $\Ym^{\mf}_{b,u,n}$ for $a_{u,n} = 1$, where the parameter is the active user position $\ysf \in \Gc_u$).

However, even by conditioning on $\ysf$, the distribution of the time-space channel matrix $\Hm_{b,u,n}$  is not fully known. For tractability, we adopt a mismatched model in which the top $D+1$ rows of $\Hm_{b,u,n}$ (see \eqref{eq:alltop_matrix}) are modeled as i.i.d. with entries $\sim \Cc\Nc(0,\sigma_{b,u,n})$ with $\sigma_{b,u,n} > 0$ representing the average received energy per tap per antenna for the hypothesized user position $\ysf$. The variance $\sigma_{b,u,n}$ can be learned offline as a function of $\ysf$ from the network geometry (details in Section \ref{subsec:radio_map}). This modeling is motivated by noticing that $\EE[\Hm_{b,u,n}] = \zerov$ regardless of the presence of \ac{los} and the exact covariance depends on the detailed local scattering environment. While knowing the average received signal energy per channel tap is possible, the a priori knowledge of the channel covariance matrix for random access users is completely unrealistic.\footnote{We want to stress here that the {\em assumed} statistics are used to derive the form of the detectors/estimators. The unavoidable mismatch between the assumed and the true channel statistics is fully taken into account in our simulations.}
Under these assumed statistics, the \ac{llr} decision metric, derived in full in Appendix~\ref{app:derivation_td-glrt}, is given in \eqref{eq:LLR_td} (at the top of the next page),
\begin{figure*}[htb!]
\begin{equation}
\Lambda^\td\big(\ysf\big)\! =\! \sum_{b\in\Vc_u}\Lambda^\td_b(\ysf)\! =\! \sum_{b\in\Vc_u}\left[\frac{\sigma_{b,u,n}\Ec_{\rm zc}}{\sigma_w(\sigma_{b,u,n}\Ec_{\rm zc}+\sigma_w)}\trace\!\Big(\big(\Ym^{\mf}_{b,u,n}\big)^\herm\Dm_{\ell_{b,u,n}(\ysf)}\Ym^{\mf}_{b,u,n}\Big) - M(D+1)\log\!\Big(1+\frac{\sigma_{b,u,n}\Ec_{\rm zc}}{\sigma_w}\Big)\right]\!.
\label{eq:LLR_td}
\end{equation}
\vspace{-3mm}
\end{figure*}
and the \ac{glrt} to detect activity is given by 
\begin{equation}
\max_{\ysf\in\Gc_u}\;\Lambda^\td(\ysf) \;\underset{\hat a_{u,n}=0}{\overset{\hat a_{u,n}=1}{\gtrless}}\; \eta_\thr^\td,
\label{eq:GLRT_td}
\end{equation}
where $\eta_\thr^\td$ is a threshold that is set to trade off the \ac{fa} and \ac{md} probabilities. The intuitive interpretation of the above decision rule is that, since the channel impulse response at each antenna of each receiving \ac{ru} is contained in a window of $D+1$ samples with an unknown cyclic shift of $\ell_{b,u,n}$ positions, the \ac{glrt} searches the windows that contain the largest received signal energy over all sets of delays corresponding to possible user positions in the grid, and then compares the maximum (over $\ysf$) signal energy with a threshold. 
In fact, it can be noticed that the metric \eqref{eq:LLR_td} consists of a weighted sum over the RUs $b \in \Vc_u$ of the received signal energies in time windows of $D+1$ samples with cyclic shifts $\ell_{b,u,n}(\ysf)$ discounted by logarithmic penalty terms that depends on the channel average strength $\sigma_{b,u,n}$. 

As a byproduct, the grid point corresponding to the maximum of \eqref{eq:GLRT_td} is itself a (coarse) position estimate of the active user $(u,n)$, assuming that codeword $(u,n)$ is detected as active. However, to limit the complexity of activity detection, the search grid $\Gc_u$ is quite coarse. For this reason, in the next section we consider applying positioning refinement only to the users detected as active.

\subsection{Active User Positioning Refinement}
\label{subsec:td-pos}

Letting $\widehat{\Acal}_u$ denote the list of codewords detected as active by the \ac{glrt} for location $\Dc_u$. For the codewords $(u,n) \in \widehat{\Acal}_u$, we propose a positioning refinement strategy using an \ac{mmle} that exploits the properties of the LoS components. 
Since the position estimation is performed separately for each individual detected codeword, we drop the indices $u,n$ and keep the \ac{ru} index $b \in \Vc_u$.

Let $\Vc_u^{\rm LoS} \subseteq \Vc_u$ denote the subsets of RUs used for detection that have \ac{los} paths to the positions of location $\Dc_u$. 
Recalling from \eqref{eq:delay_int_frac}, \eqref{eq:discrete_time_space_ch_mat} that the \ac{los} component occupies (at most) the first two non-zero rows of $\Jm^{\ell_{b}} \Hm_{n}$. We propose to isolate these two rows of the \ac{mf} output, and construct a likelihood function that exploits the explicit form of the \ac{los} component.
Let $\underline\yv_{b,1},\underline\yv_{b,2}$ denote the rows $\ell_{b}$ and $\ell_{b}+1$ of observation \eqref{eq:mf_out_td}. Under hypothesis $\Hc_1$, i.e., $(a_{u,n} =1)$ since $(u,n) \in \widehat{\Acal}_u$, and for $b \in \Vc_u^{\rm LoS}$, these rows contains the LoS components in the form
\begin{equation}
\begin{aligned}
    \underline\yv_{b,1} & = \sqrt{\Ec_{\rm zc}\beta_{b}^\los}\,\mu_{b,0}\,e^{j\phi_{b}^\los}\underline\av(\theta_{b}^\los) + \underline\wv''_{b,1}, \\
    \underline\yv_{b,2} & = \sqrt{\Ec_{\rm zc}\beta_{b}^\los}\,(1-\mu_{b,0})\,e^{j\phi_{b}^\los}\underline\av(\theta_{b}^\los) + \underline\wv''_{b,2},
\label{eq:td_pos_obs}
\end{aligned}
\end{equation}
where $\mu_{b,0} \in (0,1)$ denotes the fractional delay of the LoS path as defined below in \eqref{eq:g_sampled}, and $\underline\wv''_{b,1},\underline\wv''_{b,2}$ are the corresponding rows of the noise matrix in \eqref{eq:mf_out_td}, treated as independent with \ac{iid} components $\Cc\Nc(0,\sigma_w)$.\footnote{We want to stress again here that this is the {\em assumed} statistics used to derive the \ac{mmle}. Inevitably there is some mismatch which is fully taken into account in our simulations. The main issue here is that some level of mismatch is in any case inevitable since the exact statistics of these quantities is generally unknown to the receiver.} 
Stacking $\underline\zv_b = [\underline\yv_{b,1},\underline\yv_{b,2}]\in\CC^{1\times2M}$ of \eqref{eq:td_pos_obs}, the \ac{pdf} of this \ac{rv} is an instance of the phase-randomized model of Lemma~\ref{lem:bessel_pdf_derivation} in Appendix~\ref{app:bessel} with mean vector
\begin{equation}
\underline\uv_{b} = \sqrt{\Ec_{\rm zc}\beta_{b}^\los}\,\big[\mu_{b,0}\underline\av(\theta_{b}^\los),\;(1-\mu_{b,0})\underline\av(\theta_{b}^\los)\big].
\label{eq:td_pos_lemma_sub}
\end{equation}
After some algebra and after dropping irrelevant additive constants, the log-likelihood of $\{\underline\zv_b:b\in\Vc_u\}$ as a function of the hypothesized position $\ysf$ is given by \eqref{eq:td_ML_refined} with $w_b \eqdef \underline\av(\theta_{b}^\los)\big(\mu_{b,0}\underline\yv_{b,1}+(1-\mu_{b,0})\underline\yv_{b,2}\big)^\herm$.
\begin{figure*}[htb!]
\begin{equation}
\Lc^\td_{\mathrm{refined}}(\ysf) = \sum_{b\in\Vc_u^{\rm LoS}}\left[\log I_0\!\left(\frac{2\sqrt{\beta_{b}^\los \Ec_{\rm zc}}}{\sigma_w}\,\abs{w_b}\right) - \frac{\beta_{b}^\los \Ec_{\rm zc} M}{\sigma_w}\big(\mu_{b,0}^2+(1-\mu_{b,0})^2\big)\right].
\label{eq:td_ML_refined}
\end{equation}
\vspace{-3mm}
\end{figure*}
In \eqref{eq:td_ML_refined}, the first term has the classical form of non-coherent detection, rewarding the alignment between the array response at $\theta_{b}^\los$ and the fractional-delay-weighted combination of the two observed rows of the received signal. The dependence on $\ysf$ is (implicitly) contained in \acp{aoa} $\{\theta_{b}^\los\}$, fractional delays $\{\mu_{b,0}\}$, and path loss coefficients $\{\beta_{b}^\los\}$. Hence, the proposed \ac{mmle} can be interpreted as a joint \ac{tdoa}, \ac{aoa}, and \ac{rss} estimator. 

Since the coarse \ac{glrt} statistic $\Lambda^\td(\ysf)$ evaluated over the grid $\Gc_u$ is generally not a smooth (unimodal) surface, but it may exhibit several local maxima, a further improvement consists of performing a local refined search (i.e., maximizing $\Lc^\td_{\mathrm{refined}}(\ysf)$) around the top $K \geq 1$ maxima of $\Lambda^\td(\ysf)$ over $\Gc_u$, where $K$ is a system parameter. 
The refinement log-likelihood function in \eqref{eq:td_ML_refined} is then maximized only over a restricted fine grid $\Gc_u'$ placed around the said $K$ maxima. In this way, the complexity of the \ac{mmle} refined search is greatly reduced with respect to a full search over the entire parameter space $\ysf \!\in\! \Dc_u$, providing an effective super-resolution technique since the positioning accuracy is not limited by the signal bandwidth. 

\section{Frequency-Domain Scheme for Joint Activity Detection and User Positioning}
\label{sec:fd_pipeline}

\subsection{Frequency-space Channel Model and Statistics}
\label{subsec:freq_space_ch_stat}

Recalling \eqref{hbreve} and dropping indices $u,n$ for notation simplicity,  the $L_f\times M$ frequency-space channel matrix is defined
\begin{equation}
\begin{aligned}
    \breve\Hm_{b} &= \II^\los_b \sqrt{\beta_{b}^\los}\,e^{j\phi_{b}^\los}\,
    \vv_{b}(\ell_0,\mu_0) 
    \underline\av(\theta_{b}^\los) + \\& \sum_{p\in\Pc_b}\rho_{b,p}\sqrt{\beta_{b,p}}\,
    \vv_{b}(\ell_p,\mu_p) 
    \underline\av(\theta_{b,p}),
\label{eq:breveH}
\end{aligned}
\end{equation}
where $\vv_{b}(\ell,\mu) \eqdef \Phim^{\ell} \bv_{b}(\mu)$ and $\bv_{b}(\mu) \in \CC^{L_f}$ is the frequency-domain representation of the fractional delays due to pulse shaping and sampling, with elements given by 
\begin{equation}
[\bv_b(\mu)]_{\xi} = \mu+(1-\mu)e^{-j\frac{2\pi}{L_f}\xi}\; \text{with} \; \xi \!=\!0,\! \ldots, L_f\!-\!1.
\label{eq:frac_delay_freq}
\end{equation}
We can see that the terms in \eqref{eq:breveH} are rank-1 matrices formed by the product of the (column) frequency-domain delay steering vectors of the type $\vv_{b}(\ell,\mu)$ and the (row) \ac{ula} steering vectors of the type $\underline\av(\theta)$. 
The channel $\breve\Hm_{b}$ is a function of the fading variables (e.g., random scattering coefficients $\rho_p$, random LoS phase $\phi_{b}^\los$), conditioned on the transmitter position $\ysf$ (which fixes the \acp{aoa} and delays) and the geometry of the scatterers. In particular, the latter is the result of random placement, but fixed for a specific network, as it depends on the fine details of the coverage area geography and is generally unknown to the receiver. Conditionally on $\phi^\los$ and the scatterers' positions, each row  $\underline{\breve\hv}_b[\xi]$ of $\breve\Hm_{b}$ follows the Gaussian statistics
\begin{equation}
\underline{\breve\hv}_b[\xi]\,\big|\,\phi_{b}^\los \;\sim\; \Cc\Nc\big(\underline\mv_{b,0}[\xi]\,e^{j\phi_{b}^\los},\,\Sigmam_b[\xi]\big),
\label{eq:ch_per_ru_user_stochastic}
\end{equation}
where $\underline\mv_{b,0}[\xi]$ is the $\xi$th row of 
$\II^\los_b \sqrt{\beta_{b}^\los}
    \vv_{b}(\ell_0,\mu_0) 
    \underline\av(\theta_{b}^\los)$ in \eqref{eq:breveH} indicates the (possibly zero) \ac{los} mean vector, and the spatial covariance matrix is
\begin{equation}
\Sigmam_b[\xi] = \sum_{p\in\Pc}\beta_{b,p}\,|[\bv_{b}(\mu_p)]_{\xi}|^2\,\underline\av^\herm(\theta_{b,p})\underline\av(\theta_{b,p}).
\label{eq:cov_freq_spec_exact}
\end{equation}
The aggregate channel for a generic user over all $B$ \acp{ru} is $\underline{\breve{\hv}}[\xi] \!=\! \big[\underline{\breve{\hv}}_{1}[\xi], \!\ldots,\underline{\breve{\hv}}_{B}[\xi]\big]\!\in\! \CC^{1 \!\times\! F}$ and the blocks $\underline{\breve{\hv}}_{b}[\xi]$ are mutually statistically independent but not identically distributed.

\subsection{Offline Learned Channel Statistics}
\label{subsec:radio_map}

The covariance matrix $\Sigmam_b[\xi]$ in \eqref{eq:cov_freq_spec_exact} is generally not fully known to the receiver, since this depends on the fine details of the geometry, which in turn depends on a large number of unknown parameters. Several recent works \cite{Yapar2023Real,Zeng2024Tutorial,Jaensch2025Radio} demonstrated the ability to learn, via suitably trained deep-learning models, the \acp{lsfc} of realistic propagation environments.
In particular, a \ac{rm} yields the values expected channel strength $1/(ML_f) \sum_\xi\EE[ \|\underline{\breve{\hv}}_{b}[\xi]\|^2]$ for a given $\xsf_b$ (RU $b$ position) at each point $(\ysf \in \Rc)$ of ``grid-based" discretization of the coverage area $\mathcal{D}$ (e.g., see \cite{Yapar2023Real,Jaensch2025Radio}).
Hence, we can safely assume that such \acp{lsfc} knowledge can be learned at any position in the considered area for each \ac{ru} as a property of the scattering geometry. 

A key strength of the AMP framework is its guaranteed convergence to the asymptotic performance predicted by the \ac{se} (see \cite{Cakmak2025Joint} for details) for any Lipschitz-continuous denoising function used in the AMP iterations. The only element of the AMP algorithm where the statistics of the signal matrices $\Xm_u[\xi]$ (see \eqref{eq:y_fd_cell_free_total_subc}) play a role is in said denoising function. The {\em Bayes-optimal} denoiser takes on the form of a posterior expectation. When the signal statistics are not fully known, any assumed (possibly greatly simplified) statistics yields a valid denoiser. Of course, the validity of the theory does not imply that any choice for the assumed statistics yields good performance and it is up to the system designer to devise suitable assumptions that yield good performance.\footnote{Another line of work consists of developing message-passing algorithms or integrating \ac{em} with \ac{amp} to learn the channel priors such as \cite{Jiang2023EMAMP,Zhang2024Activity}.
In the case of multisource \ac{amp} \cite{Cakmak2025Joint}, used in this work, such algorithms have not yet been developed, and such a direction may be considered for future work.}

In this work, motivated by the learnability of \acp{rm}, 
we propose an AMP algorithm that treats the channels as zero-mean Gaussian vectors with i.i.d. segments, i.e., it assumes $\underline{\breve{\hv}}_{b}[\xi] \sim \Cc\Nc(\zerov,\bar\Sigmam_b(\ysf))$ where $\bar\Sigmam_b(\ysf)$ is diagonal and depends on the postulated transmitter position $\ysf$ only through the \acp{lsfc}. In particular, we define
\begin{align}
    \bar\Sigmam_b(\ysf) &= \bar\Sigmam_b^\los(\ysf) + \bar\Sigmam_b^\nlos(\ysf), \label{eq:tot_mis_cov}\\
    \bar\Sigmam_b^\los(\ysf) &= \II^\los_b\,\tfrac1M\trace\!\big(\beta_{b}^\los(\ysf)\underline\av^\herm(\theta_{b,0})\underline\av(\theta_{b,0})\big)\Id_M = \II^\los_b\,\beta_{b}^\los(\ysf)\Id_M, \label{eq:los_mis_cov}\\
    \bar\Sigmam_b^\nlos(\ysf) &= \bar\beta_b^\nlos(\ysf)\Id_M, \label{eq:nlos_mis_cov}
\end{align}
where, with some abuse of notation, we let channel \acp{lsfc} $\beta_{b}^\los$, $\bar\beta_{b}^\nlos$ indicate the dependency on $ \ysf$ explicitly.\footnote{In our simulation environment, these coefficients have been estimated for each given layout of scatterers and \acp{ru} by averaging over many realizations of the random scattering coefficients, thus mimicking the data-driven approaches used in \ac{rm} prediction.} 

Moreover, the per-tap strengths used in \eqref{eq:LLR_td} are directly obtained as $\sigma_{b}(\ysf) = (\beta_{b}^\los(\ysf) + \bar\beta_b^\nlos(\ysf))/(D+1)$. Learning the noise-plus-interference strength $\sigma_w\ge N_0$ used in \eqref{eq:LLR_td} and in \eqref{eq:td_ML_refined} is more challenging since this depends on the random user activity itself. In our numerical results, 
we used the lower bound $\sigma_w = N_0$ in \eqref{eq:LLR_td} and \eqref{eq:td_ML_refined}, trusting the fact that, for long ZC sequences, the cross-correlation is very small.

\subsection{AMP with Mismatched Posterior Mean Denoiser}
\label{subsec:fd-denoiser}

The multisource AMP algorithm can be found in \cite{Cakmak2025Joint} and it is omitted here for the sake of space. The key step in the AMP iteration is the non-linear denoising function $\eta_{u,t}:\CC^{F}\to\CC^{F}$, which depends on the (assumed) statistics of the $u$th signal matrix $\Xm_u[\xi]$ and on the iteration index $t$. From the AMP theory \cite{Cakmak2025Joint}, we know that, under certain technical conditions that are all verified here, the AMP output statistics at iteration $t$ converges (for large $Q$ and fixed ratios $N_u/Q$ for $u \in [U]$) to the so-called ``decoupled observation model'' in the form
\begin{equation}
    \Rm_u^{(t)}[\xi] = \Xm_u[\xi] + \Psim_u^{(t)}[\xi],   \label{eq:AMP-output}
\end{equation}
where $\Psim_u^{(t)}[\xi]$ is a zero-mean Gaussian matrix with independent components whose variances can be calculated via the multisource \ac{amp} \ac{se} for every iteration $t$ in \cite[Def. 1]{Cakmak2025Joint}, independent of $\Xm_u[\xi]$. The corresponding Bayes-optimal denoiser is the posterior mean estimation (PME) $\eta_{u,t}(\Rm) = \EE[ \Xm  | \Rm ]$, where $\Rm = \Xm + \Psim$, and $\Xm$ and $\Psim$ are mutually independent, $\Xm$ is distributed as $\Xm_u[\xi]$, and $\Psim$ is distributed as $\Psim_u^{(t)}[\xi]$. 
Since $\Xm_u[\xi]$ has independent rows (see definition above \eqref{eq:y_fd_cell_free_total_subc}), the PME splits into a per-row PME. We focus on the PME functional form, omitting indices $u, \xi$, and iteration $t$. In our case, under the assumed zero-mean Gaussian i.i.d statistics given in Section \ref{subsec:radio_map}, the denoiser takes on the same form as in \cite{Cakmak2025Joint}, given by 
\begin{equation}
    \eta(\underline{\rv}\vert \ysf) = \frac{\underline{\rv} \Gammam(\ysf)}{1+\frac{1-\lambda}{\lambda} \frac{\abs{ \tilde{\Sigmam}(\ysf) +\tilde{\Cm}}}{\abs{ \tilde{\Cm}}} {\rm e}^{-\underline{\rv}\tilde{\Cm}^{-1}\Gammam^\herm(\ysf)\underline{\rv}^\herm}} 
    \label{eq:denoiser_typeI}
\end{equation}
where $\underline{\rv}$ is any row of $\Rm$, where $\tilde{\Cm} = \blkdiag\big(\Cm_1, \ldots, \Cm_B)$ with $\Cm_b = \sigma^2_{\se,b}\Id_M$ is the covariance of the rows of $\Psim$ calculated by the \ac{se}, $\tilde{\Sigmam}(\ysf)= \blkdiag\big(\bar{\Sigmam}_{1}(\ysf), \dots, \bar{\Sigmam}_{B}(\ysf)\big)$, $\Gammam(\ysf) = (\tilde{\Sigmam}(\ysf) +\tilde{\Cm})^{-1}\tilde{\Sigmam}(\ysf)$, and $\lambda \!=\! \lambda_u\! =\! \PP(a_{u,n}\! = \!1)$ is the activity probability of the random access users in location $\Dc_u$.

It is important to recall that the denoiser depends on the postulated transmitter position $\ysf \in \Dc_u$. Hence, further averaging with respect to $\ysf$ is required. A numerically feasible approach consists of defining a grid of discrete positions $\Gc_u \! \subset \! \Dc_u$ and summing over $\ysf \!\in\! \Gc_u$. 
To limit the complexity of the scheme, we used only one point in the grid for each location $u$, namely, we choose the center of each location (marked with a red sign (+) in Fig.~\ref{fig:network_topology}.

\subsection{Activity Detection by GLRT and Positioning Refinement}
\label{subsec:fd-glrt-pos}

We consider $L_f$ parallel multisource \ac{amp} outputs $\{\Rm_u^{(T)}[\xi] : u \in [U]\}$ of \eqref{eq:AMP-output} after $T$ iterations. 
For all $n \in [N_u]$, let $\underline{\rv}_{u,n}[\xi] \eqdef [\Rm_u^{(T)}[\xi]]_{n,:}$ denote the $n$-th row corresponding to codeword $(u,n)$. 
The activity of a codeword $(u,n)$ yields the composite hypothesis test
\begin{subequations}
\label{eq:composite_fd}
\begin{align}
\Hc_0:&\quad \underline\rv_{u,n}[\xi] = \underline\psiv_{u,n}[\xi], &&\forall\xi,\\
\Hc_1:&\quad \underline\rv_{u,n}[\xi] = \underline{\breve\hv}_{u,n}[\xi] + \underline\psiv_{u,n}[\xi], &&\forall\xi .
\end{align}
\end{subequations}
where $\underline\psiv_{u,n}[\xi] \sim \Cc\Nc(\zerov, \tilde{\Cm}_u[\xi])$ and  $\tilde{\Cm}_u[\xi]$ has the block-diagonal with diagonal blocks form of the generic $\tilde{\Cm}$ defined in Section \ref{subsec:fd-denoiser}.
Problem \eqref{eq:composite_fd} is composite because hypothesis $\Hc_1$ depends on the postulated position of the $(u,n)$ user $\ysf$, which in turn determines the AoA and delay of the LoS component and the \acp{lsfc} of the LoS and NLoS components as detailed in the channel model and statistics in \eqref{eq:breveH}-\eqref{eq:cov_freq_spec_exact}.

Let $\underline{\rv}_{b,u,n} [\xi]$ denote the $\nth{b}$ $1 \times M$ block of $\underline{\rv}_{u,n} [\xi]$ and recall that under both hypotheses the blocks of $\underline{\rv}_{u,n} [\xi]$ are mutually independent for different $b \in [B]$. For each $b \in \Vc_u$, extract the $b$th blocks and stack them in $1 \times M L_f$ vectors (i.e., using all subcarriers $\xi = 0, \ldots, L_f-1$). For notational simplicity, we drop the indices $(u,n)$. Then, under $\Hc_1$ and for given $b \in \Vc_u$, the stacked vector takes on the form
\begin{equation}
    \underline{\rv}_b =
    \Big[
      \underline{\breve{\hv}}_b[0]  + \underline{\psiv}_b[0],\;
      \ldots,\;
      \underline{\breve{\hv}}_b[L_f-1] + \underline{\psiv}_b[L_f-1]
    \Big].
    \label{eq:rv_deffinal}
\end{equation}
Recalling $\breve{\Hm}_{b} \in \CC^{L_f \times M}$ of \eqref{eq:breveH}, the identity $\vec(\Am\Bm\Cm)=(\Cm^\Tran\otimes\Am)\vec(\Bm)$, using \eqref{eq:ch_per_ru_user_stochastic} and the assumed statistics of the NLoS component, and Lemma~\ref{lem:bessel_pdf_derivation} in Appendix~\ref{app:bessel}, $\underline{\rv}_b$ of \eqref{eq:rv_deffinal} has \ac{pdf} in the form \eqref{eq:lemma_result_bessel_pdf} with mean vector
\begin{equation}
    \underline{\muv}_{b,0} =\II^\los_b \sqrt{\beta_{b}^\los}    \Big(\Phim^{\ell_0}\bv_b(\mu_0)\Big)^\Tran \otimes \,\underline{\av}(\theta_{b}^\los) \in\CC^{1\times ML_f}.
    \label{eq:mean_vec_loc}
\end{equation}
Notice that, once again, the dependence on the transmitter position $\ysf \in \Dc_u$ is implicitly given through $\beta_{b}^\los$, $\ell_{b,0}$, $\mu_{b,0}$, and $\theta_{b}^\los$. The covariance matrix of $\underline{\rv}_b$ has the structure
\begin{equation}
    \Km_b = \Km_b^{\nlos} + \Km_b^{\Psi},
    \label{eq:Kb_decomp}
\end{equation}
\begin{equation}
\label{eq:block_cov}
    \Km_b\!=\!
    \begin{bmatrix}
    \Km_b^{(0,0)} & \Km_b^{(0,1)} & \cdots & \Km_b^{(0,L_f-1)} \\
    \Km_b^{(1,0)} & \Km_b^{(1,1)} & \cdots & \Km_b^{(1,L_f-1)} \\
    \vdots & \vdots & \ddots & \vdots \\
    \Km_b^{(L_f-1,0)} & \Km_b^{(L_f-1,1)} & \cdots & \Km_b^{(L_f-1,L_f-1)}
    \end{bmatrix},\!
\end{equation}
where each block $\Km_b^{(l,l')} \in \CC^{M \times M}$ is the spatial covariance between subcarriers $l$ and $l'$. After some algebra (omitted for brevity) and by applying the mixed-product property of the Kronecker product and the property $(\Phim^\ell)^* = \Phim^{-\ell}$, we get
\begin{equation}    
    \Km_b^{\nlos}
    \!=\!\sum_{p \in \Pc_b} \!\beta_{b,p}\! \underbrace{\left ( \Phim^{-\ell_p}\bv_b^*(\mu_p)\bv_b^\transp(\mu_p)\Phim^{\ell_p} \right )}_{L_f\times L_f}  \!\otimes \!
    \underbrace{\left ( \underline{\av}^\herm(\theta_{b,p})\underline{\av}(\theta_{b,p})\right )}_{M\times M}.
    \label{eq:Kb_nlos}
\end{equation}
The multisource \ac{amp} output statistics leading to \eqref{eq:composite_fd} are conditionally independent with respect to the \acp{ru} but, for each RU $b$, conditionally dependent over the subcarriers (due to the frequency-correlated channel model). However, for $L_f = D+1$ the subcarriers are widely spaced beyond a coherence bandwidth (which can be roughly quantified as $W/D$ Hz) and, in addition, the frequency correlation is a priori unknown since it depends on the fine structure of the multipath channel. Hence, we can safely disregard such correlation in the derivation of the detector. Furthermore, since the multisource \ac{amp} is applied separately to each subcarrier, the output noise covariance is block-diagonal
\begin{equation}
    \Km_b^{\Psi}= \blkdiag\Big(\Cm_b[0],\Cm_b[1],\ldots,\Cm_b[L_f-1]\Big).
    \label{eq:Kb_phi}
\end{equation}
Hence, the resulting approximated covariance matrix is 
block-diagonal 
\begin{equation}
\label{eq:block_diag_cov}    \tilde{\Km}_b=\blkdiag\Big(\tilde{\Km}_b[0],\tilde{\Km}_b[1],\ldots,\tilde{\Km}_b[L_f-1]\Big),
\end{equation}
with blocks $\tilde{\Km}_b[\xi] = \Sigmam_b[\xi] + \Cm_b[\xi]$. This yields the PDF 
\begin{equation}
    p\big(\underline{\rv}_b|a \!=\! 1, \ysf \big)\!=\!\frac{e^{-\underline{\rv}_b\tilde{\Km}_b^{-1}\underline{\rv}_b^\herm -\underline{\muv}_{b,0}\tilde{\Km}_b^{-1}\underline{\muv}_{b,0}^\herm}}{\pi^{ML_f}|\tilde{\Km}_b|}  I_0\!\Big(2\big|\underline{\muv}_{b,0}\tilde{\Km}_b^{-1}\underline{\rv}_b^\herm\big|\Big).
    \label{eq:marginal_pdf_loc}
\end{equation}
Due to the limited knowledge of the channel statistics already discussed in Section \ref{subsec:radio_map} and similar to the denoiser discussion, in our results we have adopted the mismatched covariance $\bar{\Km}_b \in \CC ^{L_f M \times L_f M}$ in block diagonal form with diagonal blocks $\bar{\Km}_b[\xi] = \bar{\Sigmam}_b + \Cm_b[\xi]$ with the ``nominal" diagonal covariance $\bar{\Sigmam}_{b}$. Under $\Hc_0$, we have $\underline{\rv}_{b} \sim \Cc\Nc(\zerov, \Km_b^{\Psi})$ where $\Km_b^{\Psi}$ is computed using \eqref{eq:Kb_phi}.

Hence, inserting back the indices $(u,n)$ and letting\footnote{The explicit form of $\Lambda^\fd_{b,u,n}\big(\underline\rv_{b,u,n}\big)$ can be obtained after some algebra from the two conditional distributions discussed above, and it is omitted for the sake of space.} 
$\Lambda^\fd_{b,u,n}\big(\underline\rv_{b,u,n}\big) =  \log \frac{p\big(\underline\rv_{b,u,n}\mid a=1\big)}{p\big(\underline\rv_{b,u,n}\mid a=0\big)}$ the resulting (slightly) mismatched log-likelihood ratio for the \ac{glrt} takes on the additive form
\begin{equation}
\Lambda^\fd(\ysf) = \sum_{b\in\Vc_u} \Lambda^\fd_{b,u,n}\big(\underline\rv_{b,u,n}\big).
\label{eq:LLRs_all_fd}
\end{equation}

Eventually, the frequency-domain \ac{glrt} takes on the form 
\begin{equation}
\max_{\ysf\in\Gc_u}\;\Lambda^\fd(\ysf) \;\underset{\hat a_{u,n}=0}{\overset{\hat a_{u,n}=1}{\gtrless}}\; \eta_\thr^\fd,
\label{eq:GLRT_fd}
\end{equation}
where the grid $\Gc_u$ is the same as already introduced for the time-domain GLRT scheme.

In the same spirit as what was done for the time-domain approach, we also propose here a positioning refinement for the active codewords. 
In particular, for all $(u,n) \in \widehat{\Acal}_u$, the \ac{mmle} \eqref{eq:LLRs_all_fd} is restricted to a subset $\Vc_u^{\rm LoS}$ of \acp{ru} for which \ac{los} is likely (analogous to the time-domain case) and it is re-evaluated on a restricted fine grid $\Gc_u'$ placed around the $K$ coarse maxima produced by the \ac{glrt}-based estimate. 

\section{Numerical Results and Discussion}
\label{sec:Sim_Res_Disc}

\subsection{Simulation Setup and System Configuration}
\label{subsec:sim_sys_config}

Since in a user-centric \ac{cf} network an uplink signal is received at multiple \acp{ru}, the open-loop power control based on the strength of a downlink beacon from a single serving base station, as specified in the 3GPP RACH \cite{ETSI138} is not possible. 
Hence, we assume that the random access users transmit their codewords at fixed power $P_\mathrm{tx}$, resulting in the {\em reference} transmit \ac{snr}
\begin{equation}
  \SNR_\mathrm{tx} \eqdef \frac{P_\mathrm{tx}}{W N_0} = \frac{\Ec_s}{N_0},
  \label{eq:snr_def}
\end{equation}
where $\Ec_s$ is the already introduced average transmit energy per chip. This is a deterministic quantity for constant modulus sequences such as the considered \ac{zc} codebook, while for randomly generated codebooks with \ac{iid} complex Gaussian components used in the AMP-based frequency-domain scheme~\cite{Cakmak2025Joint}, it is intended in the average sense, i.e.,  $\Ec_s=\EE[|s|^2]$. Since $W$ is the same, definition \eqref{eq:snr_def} is consistent for both schemes.

In the results of this section, we consider the layout in Fig.~\ref{fig:network_topology} with the main simulation parameters summarized in Table \ref{tab:simparams}. The random access users' transmit power $P_\mathrm{tx}$ is determined by fixing a reference receive SNR $\SNR_\refsub$ for pure \ac{los} propagation (no multipath) and isotropic transmit and receive antennas (no antenna arrays), for a user placed at a hexagon center and its nearest \ac{ru}. This yields $\SNR_\mathrm{tx}=\SNR_\refsub/\beta_\refsub$ where $\beta_\refsub$ is the \ac{lsfc} of the reference path. 

\begin{table}[t]
\centering
\caption{Simulation Parameters}
\label{tab:simparams}
\begin{tabular}{ll}
\hline
Parameter & Value \\
\hline
Number of \acp{ru}, $B$ & 36 (12 sites $\times$ 3 sectors) \\
Antennas per \ac{ru}, $M$ & 8 \\
Number of locations, $U$ & 7 \\
Hexagon radius & 100~m \\
TD \ac{zc} sequence length, $L_t$ & 4591 \\
TD subcodebook size, $N_u^\td$ & 655 \\
TD cyclic prefix, $L_\cp^\td$ & 16 chips \\
FD active subcarriers, $L_f$ & 16 \\
FD subcodebook size, $N_u^\fd$ & 655 \\
FD \ac{ofdm} pilot symbols, $Q$ & 144 \\
FD cyclic prefix, $L_\cp^\fd$ & 16 chips \\
System bandwidth, $W$ & 20~MHz \\
Carrier frequency, $f_c$ & 3.5~GHz \\
\ac{nlos} scatterers per location & 35 \\
Scatterer cross-section, $\sigma_s^2$ & $-5$~dB \\
\ac{nlos} truncation radius $d^{\rm scat}_{\Srm,\Urm},d^{\rm scat}_{\Rrm,\Srm} $ & 110~m \\
\ac{los} truncation radius $d^{\rm LoS}_{\Rrm,\Urm}$ & 200.1~m \\
AMP iterations, $T$ & 20 \\
Noise power spectral density, $N_0$ & \unit[-174]{dBm/Hz}\\
\hline
\end{tabular}
\end{table}

We use the pathloss function for \ac{los} given by $\mathrm{PL}^\los(d) = \Big(\frac{\csf}{4\pi f_c d} \Big)^2$ where $f_c$ is the carrier frequency, $d$ is the distance, and the pathloss exponent is equal to 2 since we consider only \ac{los} and specular scatterers. For the \ac{nlos}, the pathloss follows a similar formula, multiplied by an attenuation factor $\sigma_s^2$ to model the scatterer cross-section. In these results, for each Monte Carlo realization, we generate active user positions, the random selection of their codewords (without preamble collisions), the set of \ac{los}/specular paths, and the channel coefficients according to the model in Section \ref{sec:sys_ch_model}. Each generation of the active user and scatters positions is averaged over 40 independent realizations of the channel coefficients.

The detection performance is evaluated in terms of \ac{md} and \ac{fa} probabilities \cite{Poor2013Introduction,Cakmak2025Joint}. The localization performance is quantified by the Euclidean distance $\|\hat{\ysf}_k-\ysf_k\|$ (i.e., the absolute position error) restricted to the true-positive set, i.e., the active users detected as active \cite{Gkiouzepi2024Joint,Gkiouzepi2025Joint}. 
Notice that only the positioning of true positive active users is relevant, because \ac{fa} events do not correspond to any real user, and \ac{md} events are active users that are missed by the detector and therefore ignored by the system. These users do not receive an ACK in the downlink and will try again to access the \ac{rach} after some time-out. 

\begin{figure*}[htb!]
  \centering
  \subfloat[RU 1: \acp{lsfc} for \ac{los} path]{\includegraphics[width=0.33\linewidth]{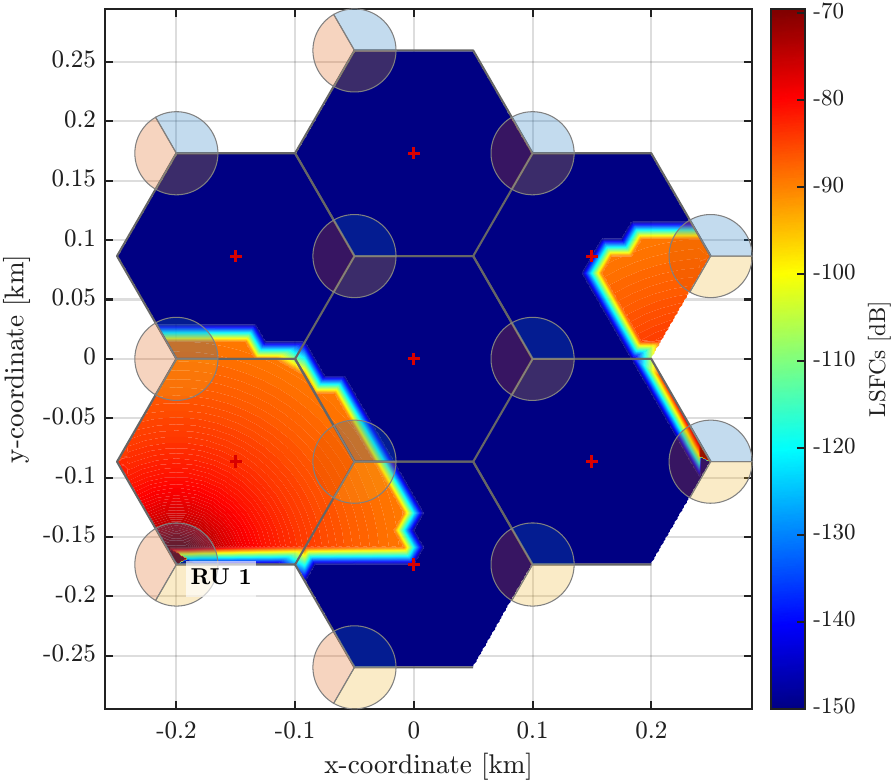}\label{fig:ru1_los}}
  \hfil
  \subfloat[RU 1: \acp{lsfc} for \ac{nlos} paths]{\includegraphics[width=0.33\linewidth]{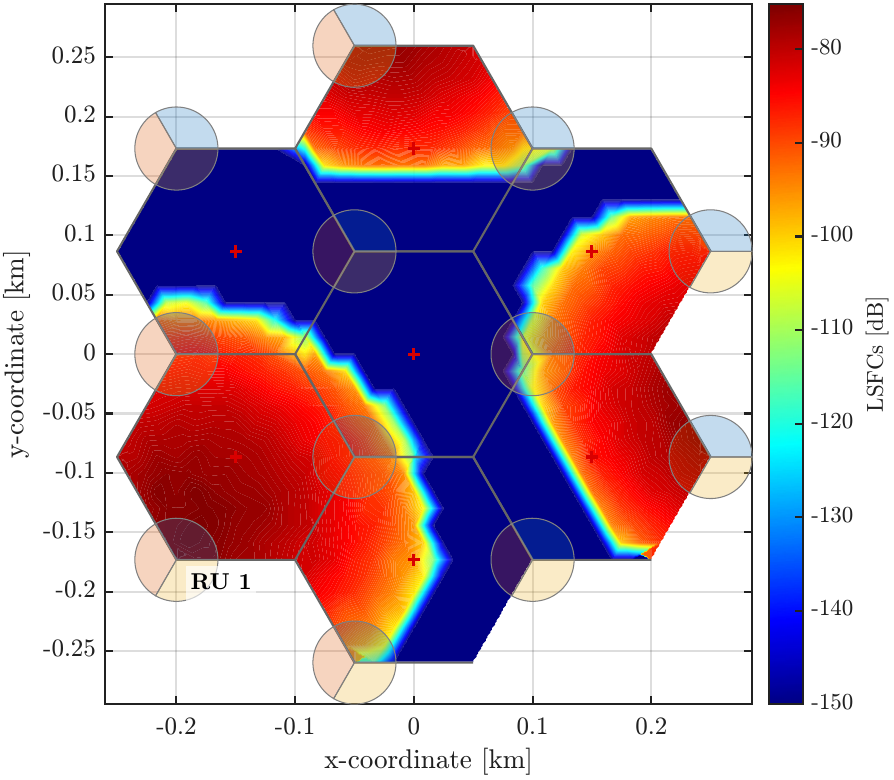}\label{fig:ru1_nlos}}
  \hfil
  \subfloat[RU 18: \acp{lsfc} for both \ac{los} and \ac{nlos} paths]{\includegraphics[width=0.33\linewidth]{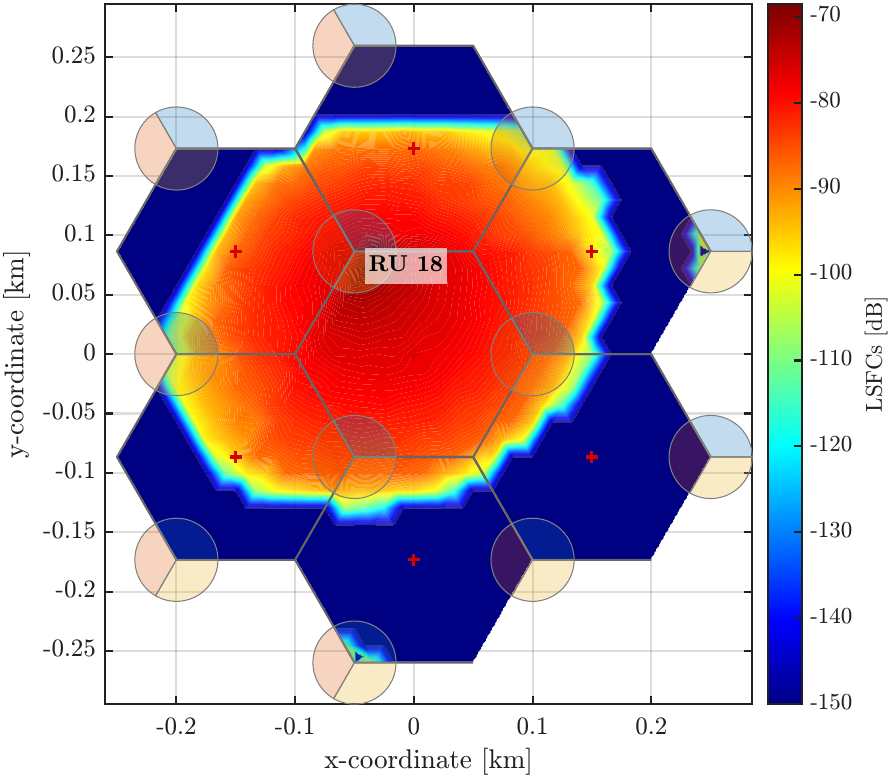}\label{fig:ru18_tot}}
  \caption{Learned (offline) spatial distribution radio maps of \acp{lsfc} across the used network: (a) and (b) the \ac{los} $\beta_{b}^\los(\ysf)$ in \eqref{eq:los_mis_cov} and \ac{nlos} $\bar\beta_{b}^\nlos(\ysf)$ in \eqref{eq:nlos_mis_cov} propagation paths seen by RU 1, and (c) the total \acp{lsfc} $\beta_{b}^\los(\ysf) + \bar\beta_{b}^\nlos(\ysf)$ in \eqref{eq:tot_mis_cov} for RU 18.}
  \label{fig:lsfc_radio_maps}
\end{figure*}
\subsection{Dimensioning the search grids for GLRT detection and position refinement}
\label{subsec:grid_pos}

For two candidate positions $\ysf,\ysf' \in \Dc_u$ to be distinguishable through the delay to \ac{ru} $b$ we need $\big|\|\ysf-\ysf_b\|-\|\ysf'-\ysf_b\|\big|W/\csf\ge1$. Then, under the (worst-case) assumption that $\ysf,\ysf',\ysf_b$ are co-linear, this yields the distance resolution limit $\|\ysf-\ysf'\|\ge\csf/W$, which we have used to lay out the \ac{glrt} search grid $\Gc_u$ as a hexagonal lattice of minimum spacing $\Delta\approx\csf/W$. For a hexagonal lattice with point spacing $\Delta$, the maximum absolute quantization error is $\Delta / \sqrt{3}$.

For the proposed \ac{glrt} approaches with positioning refinement, the coarse search is performed over $\Gc_u$ with a spacing of $\Delta_{\Gc_u} = \unit[50]{m}$ and consists of $7$ test points per location (the center coordinate, marked with a red sign (+) in Fig.~\ref{fig:network_topology}, and six symmetric points on a ring of radius $\unit[50]{m}$), yielding a maximum coarse quantization error of $50/\sqrt{3} \approx \unit[28.9]{m}$. Then, refinement grid $\Gc_u'$ is constructed with a spacing of $\Delta_{\Gc_u'} = \csf/W = \unit[15]{m}$ (for $W = \unit[20]{MHz}$), yielding maximum quantization error of $15/\sqrt{3} \approx \unit[8.7]{m}$. Notice that to cover the entire location ($\unit[100]{m}$ cell per location) at this fine resolution would require six concentric rings totaling 127 grid points. 
Hence, a brute-force search over the fine uniform grid requires $127 \times N_u$ evaluations (127 points per location for each possible codeword), resulting in a prohibitive complexity. 
In contrast, the proposed hierarchical grid scheme requires only $7 \times N_u$ evaluations for the GLRT activity detection. This is followed by $K \times |\Gc_u'|$ ($|\Gc_u'| = 27$ points around its top-$K$ coarse maxima) fine grid evaluations for each actively detected codeword, where $|\widehat{\Acal}_u| \ll N_u$ due to the sporadic random access activity.

\subsection{Offline Learned Statistical Parameters}
\label{subsec:sim_lsfc_rm_sigmaTD}

For each candidate position on a grid, the $M\times M$ per-\ac{ru} channel covariance matrices are precomputed offline on the basis of independent scatterers drops and random coefficient generations. 
The association of each grid point to these channel covariance matrices forms a generalized \ac{rm} (aka, a \ac{ckm}), 
as explained in Section \ref{subsec:radio_map}, using $100$ independent realizations. Fig.~\ref{fig:lsfc_radio_maps} shows a \ac{rm}-style visualization of the resulting \acp{lsfc}.  

\subsection{Activity Detection Performance}

\begin{figure}[htb!]
  \centerline{\includegraphics[width=0.9\linewidth]{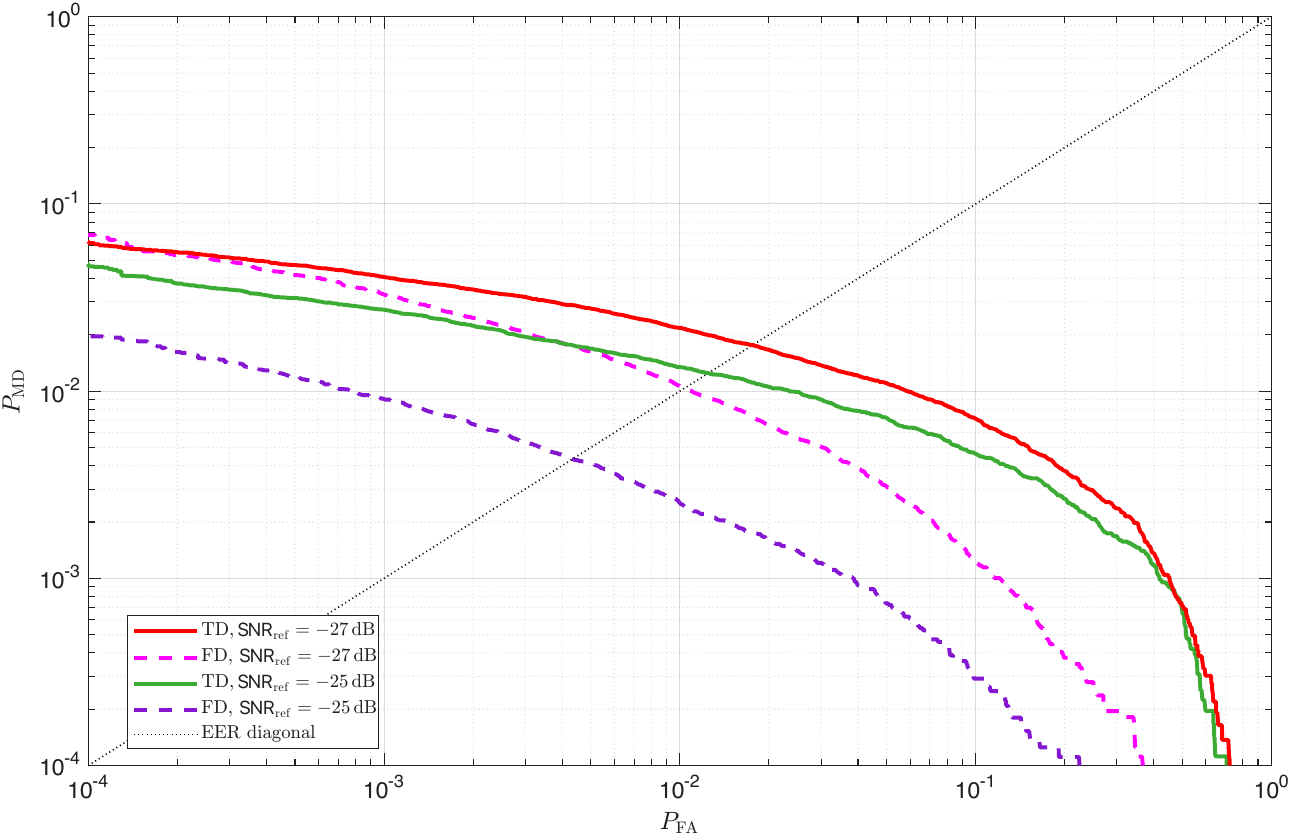}}
  \caption{MD vs. FA operating curves of the frequency-domain and the time-domain schemes for an average total number of active users equal to 600 and GLRT with $\Vc_u = \Vc_u^{\rm LoS}$.}
  \label{fig:pfa_pmd_td_fd_vsSNR}
\end{figure}

Fig.~\ref{fig:pfa_pmd_td_fd_vsSNR} shows the operating boundary of the GLRT activity detectors (i.e., the \ac{md} vs. \ac{fa} curve \cite{Poor2013Introduction}) at a fixed random activity load of (on average) 600 active users in the network, and two $\SNR_\refsub$ values ($-27$, $-25\,$dB). 
For both schemes, the corresponding GLRT is performed by using the RU set $\Vc_u = \Vc_u^{\rm LoS}$ containing the three RUs in LoS of location $\Dc_u$.\footnote{For both schemes, restricting the detection to $\Vc_u^{\rm LoS}$ yields better performance than using all the \acp{ru}.  We interpret this fact as an effect of the mismatch in the likelihood ratios, due to the discrepancy of the (known) assumed statistics and the (unknown) true statistics.} 
The \ac{eer} {\em diagonal} indicates the operating points at which the FA and MD probabilities are equal (the common value is referred to as \ac{eer} in the following).
We notice the superiority of the novel frequency-domain scheme with respect to the ``legacy-inspired'' time-domain scheme.

\begin{figure*}[htb!]
  \centering
  \subfloat[Frequency domain approach]{\includegraphics[width=0.4\linewidth]{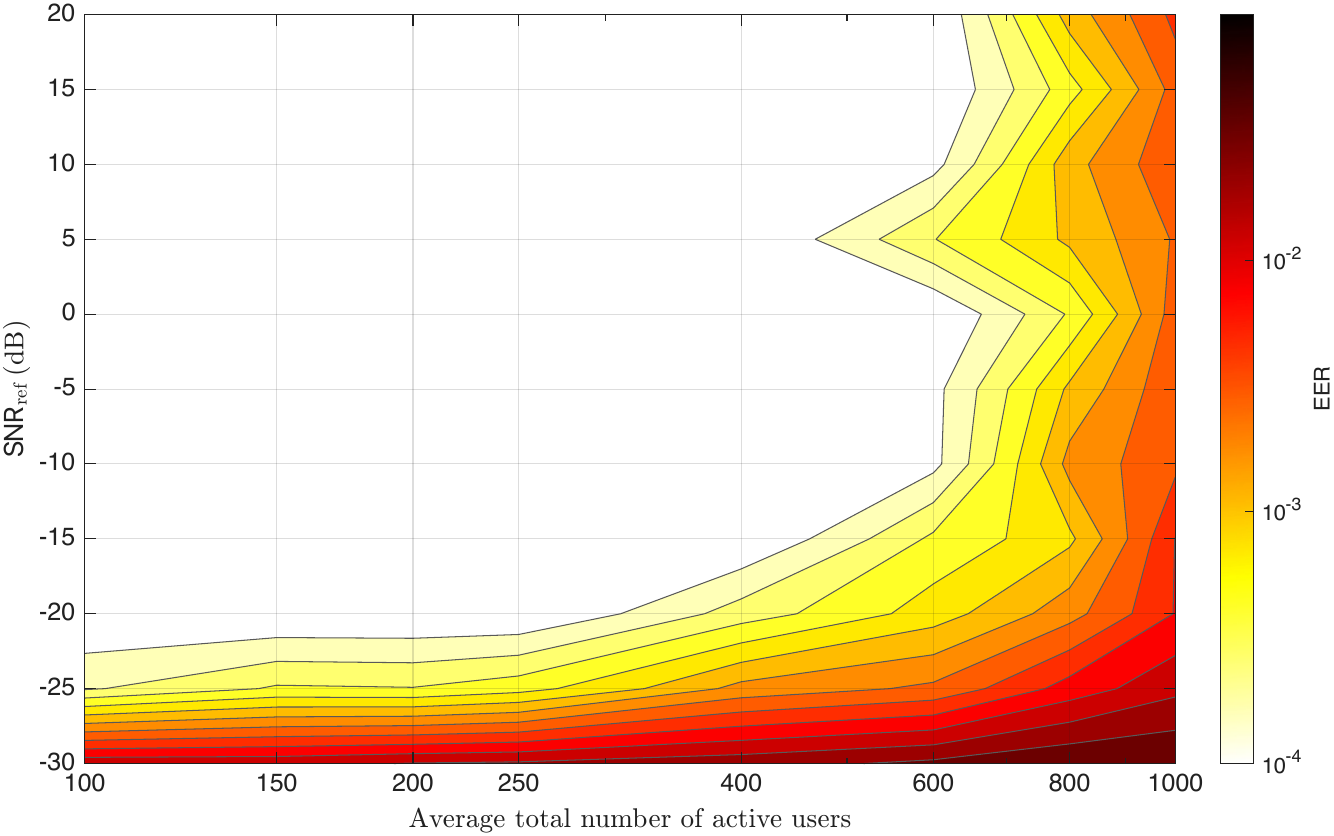}\label{fig:ad_fd_allSNR_eer}}
  \hfil 
  \subfloat[Time domain activity approach]{\includegraphics[width=0.4\linewidth]{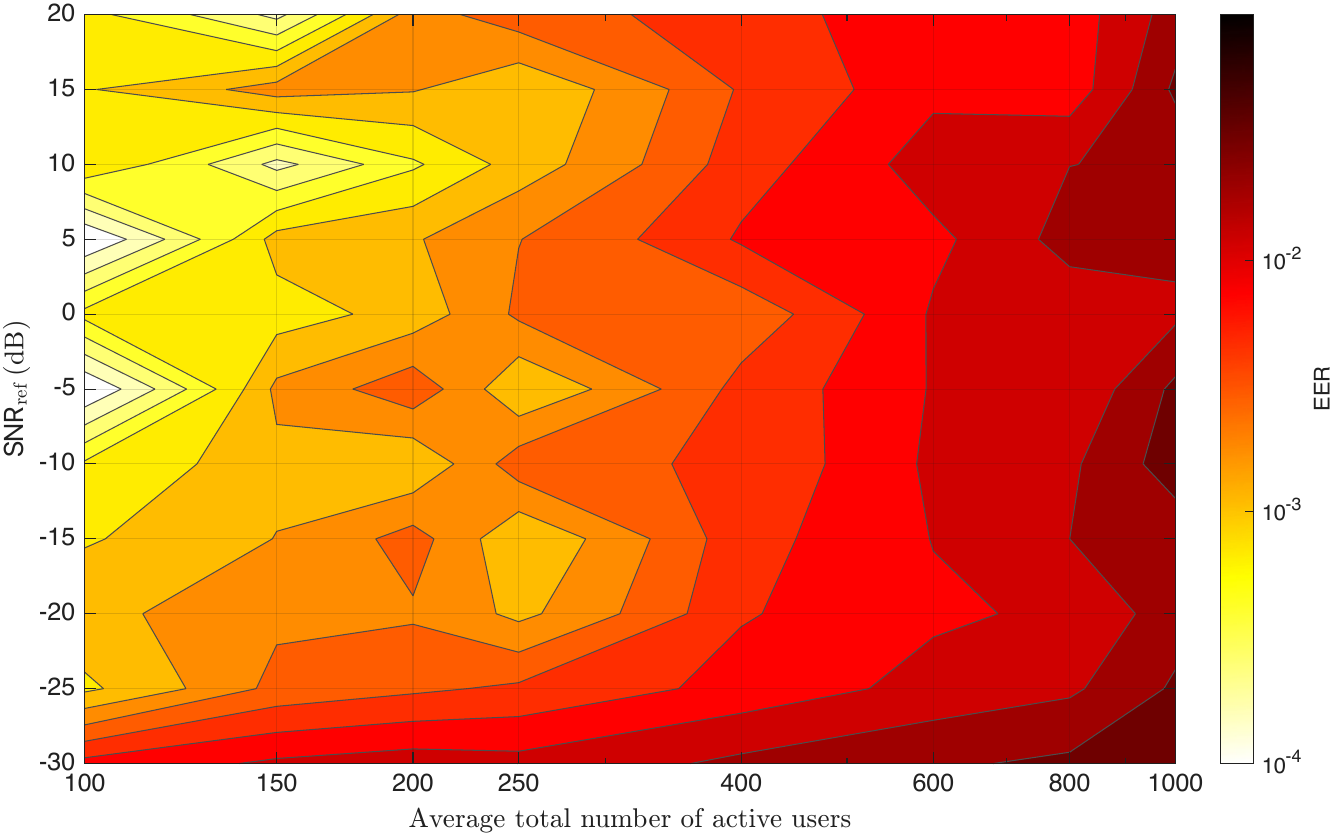}\label{fig:ad_td_allSNR_eer}}
  \caption{Performance comparison of frequency-domain and time-domain schemes for activity detection, expressed as \ac{eer} values, versus $\SNR_\refsub$ and average total number of active users.}
  \label{fig:tdvsfd_users_snr}
\end{figure*}

Fig.~\ref{fig:tdvsfd_users_snr} shows the performance of the frequency-domain and time-domain schemes for activity detection for different values of $\SNR_\refsub$ and an increasing random activity {\em load}. In particular, we consider the \ac{eer} operating point and present \ac{eer} values for each pair of (load, $\SNR_\refsub$) in the form of heat-maps in Figs.~\ref{fig:ad_fd_allSNR_eer} and \ref{fig:ad_td_allSNR_eer}, where $\SNR_\refsub$ ranges between -30 dB and 20 dB and the average total number of active users ranges from 100 to 1000. 
The superiority of the proposed \ac{amp}-based frequency-domain scheme is evidenced by the much larger region of (load, $\SNR_\refsub$) values for which the frequency-domain scheme achieves very low \ac{eer} (in fact, the white area in Fig.~\ref{fig:ad_fd_allSNR_eer} corresponds to values for which we could not count any error event, i.e., the \ac{eer} in this region is expected to be less than $10^{-4}$).

\subsection{Localization}
\begin{figure*}[htb!]
  \subfloat[Frequency-domain scheme]{\includegraphics[width=0.4\linewidth]{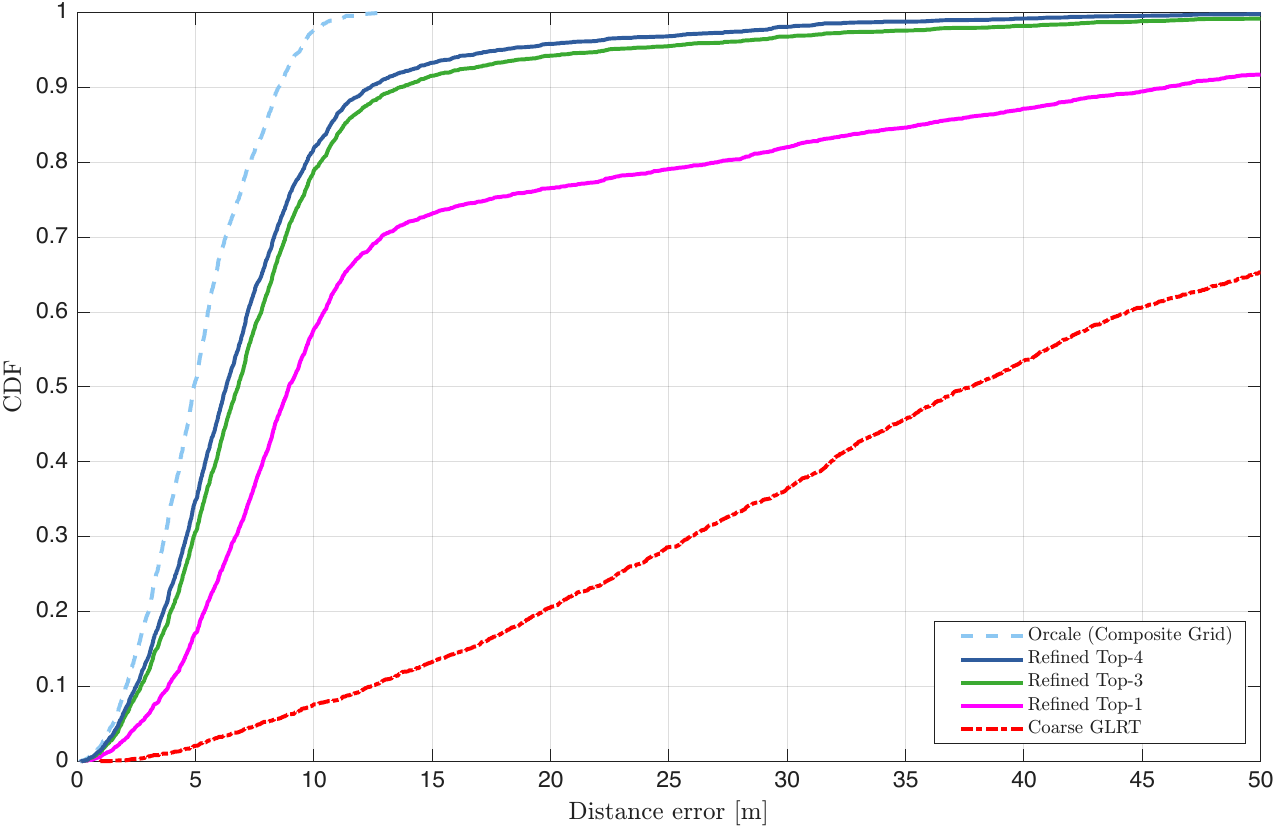}\label{fig:cdf_loc_fd}}
  \hfil
  \subfloat[Time domain scheme]{\includegraphics[width=0.4\linewidth]{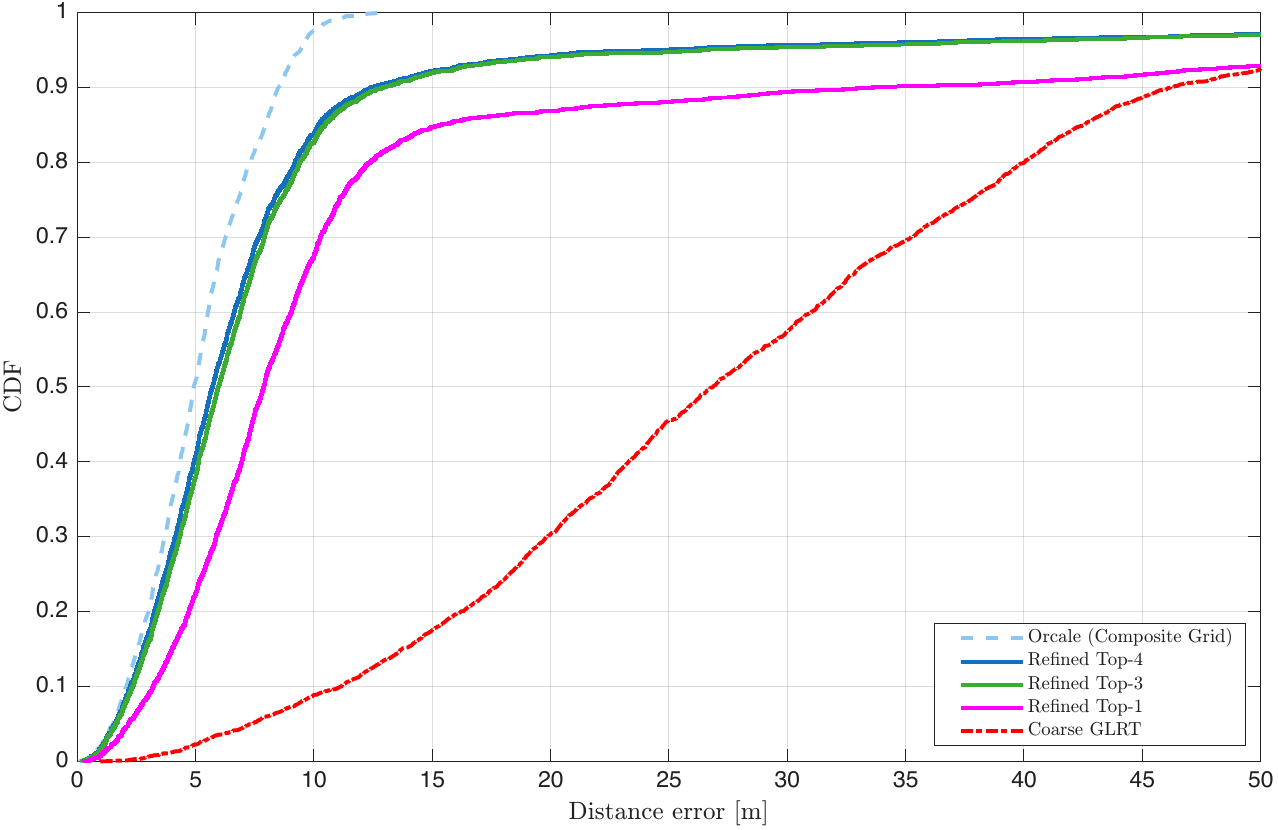}\label{fig:cdf_loc_td}}
  \caption{Localization error \ac{cdf} for the considered schemes compared with the oracle benchmark (grid discretization error) at $\SNR_\refsub = \unit[10]{dB}$ and an average total number of active users equal to 300.}
  \label{fig:td_fd_loc_eachalone}
\end{figure*}

Fig.~\ref{fig:td_fd_loc_eachalone} shows the performance of the proposed localization schemes in terms of the CDF of the absolute localization error (Euclidean distance) for the coarse localization from the GLRT activity detection and for the refinement with fine search grid $\Gc_u'$ placed around the top-1, top-3, and top-4 GLRT maxima. While the GLRT detection operating curves of Fig.~\ref{fig:pfa_pmd_td_fd_vsSNR} are given for very low $\SNR_\refsub$ to be able to count FA and MD events, here we consider the more realistic regime $\SNR_\refsub = 10$dB, corresponding to $\beta_\refsub = \unit[-83.2]{dB}$ and $P_{\rm tx} = \unit[-7.66]{dBm}$, at which the FA and MD probabilities are so small that true positives essentially coincide with the actual active users. 
For comparison, we also show the CDF of an {\em oracle} that approximates the true user position by first quantizing it on the coarse grid $\Gc_u$ and then refining the estimate on the fine grid $\Gc_u'$ centered on the coarsely-quantized point. 
The hierarchical grid-based search mechanism of the oracle is the same as for our proposed methods, but instead of evaluating the proposed likelihood function, the oracle compares directly with the true position. Hence, the oracle error CDF represents the effect of the grid quantization error. 

The localization results are obtained by choosing $\Vc_u = \Vc_u^{\rm LoS}$, for which the existence of 3 RUs with LoS propagation to all points of the corresponding location $\Dc_u$ is guaranteed. 
The existence of a substantial number of RUs with LoS propagation to any given position is one advantage of the distributed and dense user-centric \ac{cf} architecture, with respect to concentrating the antennas in a much smaller number of massive MIMO cell-sites. 

We notice that both the time-domain and the frequency-domain schemes achieve near-oracle localization performance for the true-positive users. Both schemes achieve excellent performance and prove that accurate positioning information for random access users can be extracted directly from the RACH slot, which is very attractive for massive random access in user-centric \ac{cf} networks.

\section{Conclusion}
\label{sec:conclusion}

We studied \ac{ura} for the \ac{rach} functionality in \ac{cf} user-centric networks, in a spatially consistent channel model including \ac{los} and \ac{nlos} propagation paths.
Two schemes were considered: a ``legacy-inspired'' time-domain scheme based on a \ac{zc} codebook and \ac{mf} \ac{glrt} detection, and a frequency-domain multisource \ac{amp} scheme with a \ac{glrt}-based detection.

For both schemes, the detected active users are positioned first using the \ac{glrt} as a coarse stage and then refined using an approximate \ac{mmle} method, which implicitly makes use of the dependency of delays (\ac{tdoa}), \ac{aoa}, and \ac{rss} on the user position, and naturally fuses observations across multiple \acp{ru}. The novel frequency-domain scheme dominates the time-domain one in terms of activity detection, expressed using \ac{md} and \ac{fa} probabilities. After \ac{mmle} refinement, both schemes reach close to the oracle bound (median absolute error $\approx 7$m for true-positive detected users).
The very low FA and MD probabilities even in the case of a large number of random access users, and the excellent localization performance directly from the RACH slot place the proposed AMP-based frequency-domain approach as a very suitable candidate for high-performance low-latency massive random access and enable fast user-centric cluster formation in \ac{cf} user-centric wireless networks.

\appendices
\counterwithin*{equation}{section}
\renewcommand\theequation{\thesection.\arabic{equation}}

\section{Preliminaries}
\label{app:circ_dft}

In this section, we refer to a generic block length $L$. This can be particularized to $L = L_t$ or $L = L_f$ depending on the context in the paper.  The $L\times L$ cyclic downshift matrix is defined as
\begin{equation}
\label{eq:cyc_shift}
    \Jm = \left[\begin{array}{cc}\zerov_{L-1}^\transp & 1\\ \Id_{L-1} & \zerov_{L-1}\end{array}\right],
\end{equation}
such that $\Jm\sv$ is the cyclic downshift by one position of a column vector $\sv$. It follows that $\Jm^\ell$ cyclically shifts $\sv$ by $\ell$ positions. Moreover, $\big(\Jm^{\ell}\big)^\herm = \Jm^{-\ell} = (\Jm^{\ell})^{-1}$. We introduce also the unitary \ac{dft} matrix $\Fm \in \CC^{L \times L}$ with entries $[\Fm]_{k,m} = \frac{1}{\sqrt L}\omega^{k m}$ for $k, m = 0, \ldots, L-1$, and the linear-phase shift matrix $\Phim \eqdef \mathrm{diag}(1,\omega,\ldots,\omega^{L-1})$ with $\omega \eqdef {\rm e}^{-j2\pi/L}$. The circulant matrix built on $\sv$ is 
\begin{equation}
\label{eq:cir_mat}
    \Circ(\sv)=[\sv,\Jm\sv,\ldots,\Jm^{L-1}\sv]=\sum_{\ell=0}^{L-1} [\sv]_\ell \Jm^\ell,
\end{equation}
and can also be expressed using the \ac{dft} matrix as
\begin{equation}
\label{eq:cir_mat_dft}
    \Circ(\sv)= \Fm^\herm\diag(\check\sv)\Fm, \,\text{with} \,\, \check\sv=\sqrt{L}\Fm\sv.
\end{equation}
\begin{proper}\label{prop:circ_conv_dft_hadamard} 
The cyclic convolution of a sequence $\sv\in\CC^L$ with a sequence $\xv\in\CC^L$ $(\check\xv=\sqrt{L}\Fm\xv)$ is given by
\begin{equation}
    \sv\circledast\xv = \Circ(\sv)\xv = \frac{1}{\sqrt L}\Fm^\herm\big(\check\sv\odot\check\xv\big),
\label{eq:circ_conv_dft_hadamard}
\end{equation}
\end{proper}
\begin{proper}\label{prop:conv_delayedJ}
The cyclic convolution of $\sv$ with a delayed version of $\xv$ satisfies
\begin{equation}
    \sv \circledast (\Jm^\ell \xv) =(\Jm^{\ell} \sv) \circledast \xv = \Jm^\ell(\sv\circledast\xv) = \Circ(\sv) \Jm^{\ell} \xv.
\label{eq:conv_delayedJ}
\end{equation}
\end{proper}
\begin{proper}\label{prop:F_Phi_J}
A cyclic shift in the discrete-time domain corresponds to a linear phase rotation across the \ac{dft} (discrete) frequency domain, i.e.,
$\Fm\Jm^\ell=\Phim^\ell\Fm$.
\end{proper}

\section{Derivation of Time-Domain LLR}
\label{app:derivation_td-glrt}

This appendix derives \eqref{eq:LLR_td}. We first fix a hypothesized position $\ysf$ (hence $\ell=\ell_{b,u,n}(\ysf)$ and $\sigma=\sigma_{b,u,n}(\ysf)$) and drop the $(b,u,n)$ indices for simplicity of notation. The $L_t\times M$ matrix $\Ym^\mf$ of \eqref{eq:mf_out_td} has independent entries, and under hypothesis $\Hc_1(\ysf)$, rows in the window $\{\ell,\ldots,\ell+D\}$ are \ac{iid}\ $\Cc\Nc(0,\sigma \Ec_{\rm zc}+\sigma_w)$ per entry and rows outside are \ac{iid}\ $\Cc\Nc(0,\sigma_w)$. Under $\Hc_0$ every entry is \ac{iid}\ $\Cc\Nc(0,\sigma_w)$. Writing the negative log-likelihood function and dropping additive constants independent of the hypothesis and of $\ysf$, we get 
\begin{equation}
    \begin{aligned}
    &\Lc_1(\ysf)= \frac{\trace\big((\Ym^\mf)^\herm\Dm_\ell\Ym^\mf\big)}{\sigma \Ec_{\rm zc}+\sigma_w} \!+\! \frac{\trace\big((\Ym^\mf)^\herm(\Id_{L_t}-\Dm_\ell)\Ym^\mf\big)}{\sigma_w} \\&+ M(D\!+\!1)\log(\sigma \Ec_{\rm zc}+\sigma_w) \!+\! M(L_t-D-1)\log\sigma_w,
    \end{aligned}
\label{eq:L1_appendix}
\end{equation}
\begin{equation}
\Lc_0 = \frac{\trace\big((\Ym^\mf)^\herm\Ym^\mf\big)}{\sigma_w} + ML_t\log\sigma_w,
\label{eq:L0_appendix}
\end{equation}
where $\Dm_\ell \eqdef \Jm^\ell\left[\begin{smallmatrix}\Id_{D+1}&\zerov\\\zerov&\zerov\end{smallmatrix}\right](\Jm^\ell)^\transp$ selects (and wraps around, modulo $L_t$) the $D+1$ diagonal entries between positions $\ell$ and $\ell+D$. Using $\trace\big((\Ym^\mf)^\herm\Ym^\mf\big)=\trace\big((\Ym^\mf)^\herm\Dm_\ell\Ym^\mf\big)+\trace\big((\Ym^\mf)^\herm(\Id_{L_t}-\Dm_\ell)\Ym^\mf\big)$, the \ac{llr} $\Lambda^\td_b(\ysf)=\Lc_0-\Lc_1(\ysf)$ takes on the form of the terms in the summation in \eqref{eq:LLR_td}.

\section{The PDF of zero-mean Gaussian vector plus a deterministic vector with random phase}
\label{app:bessel}
\begin{lem}
\label{lem:bessel_pdf_derivation}
Let $\underline{\xv} \in\CC^{1\times M}$ be a row random 
vector in the form $\underline{\xv} = \underline\muv e^{j\phi} + \underline{\gv}$ where $\underline\muv$ is deterministic, 
$\phi\sim\Uc[0,2\pi)$, and $\underline\gv \sim \Cc\Nc(\zerov, \Sigmam)$, such that, conditionally on $\phi$,  
\begin{equation}
\underline{\xv} |\phi \;\sim\; \Cc\Nc\big(\underline\muv\,{\rm e}^{j\phi},\,\Sigmam\big).
\label{eq:lemma_conditional_RV}
\end{equation}
Then the marginal (phase-averaged) \ac{pdf} of $\underline{\xv}$ is
\begin{equation}
p(\underline{\xv}) = \frac{{\rm e}^{-\underline{\xv}\Sigmam^{-1}\underline{\xv}^\herm - \underline\muv\Sigmam^{-1}\underline\muv^\herm}}{\pi^M \abs{\Sigmam}}I_0\big(2\abs{z}\big),
\label{eq:lemma_result_bessel_pdf}
\end{equation}
where $z = \underline\muv\Sigmam^{-1}\underline\xv^\herm$, $I_0(x)\eqdef \tfrac1{2\pi}\int_0^{2\pi}{\rm e}^{x\cos\varphi}\mathrm d\varphi$ is the modified Bessel function of the first kind and order zero. Notice that when $\underline\muv=\zerov$, \eqref{eq:lemma_result_bessel_pdf} reduces to the PDF of the Gaussian distribution $\Cc\Nc(\underline{\zerov},\Sigmam)$, 
since $I_0(0)=1$. \hfill $\square$
\end{lem}
The proof of this lemma is straightforward, and it is omitted for brevity. The result is given explicitly here since it is often used in this paper. 

\bibliography{abbrev,references}
\bibliographystyle{IEEEtran}

\end{document}

%% file: macros.tex
\usepackage{amsfonts}
\usepackage{amsmath}
\usepackage{amssymb} 
\usepackage{array}
\usepackage{bm} 
\usepackage{bbm}
\usepackage{color} 
\usepackage{cases} 
\usepackage{cite}
\usepackage{dsfont}
\usepackage[pdftex]{graphicx}
\DeclareGraphicsExtensions{.eps,.pdf,.png,.jpg,.gif,.jpeg,.pstex}
\usepackage{hyperref}
\hypersetup{
    unicode=false,          
    pdftoolbar=true,        
    pdfmenubar=true,        
    pdffitwindow=false,     
    pdfstartview={FitH},    
    pdfnewwindow=true,      
    colorlinks=true,       
    linkcolor=blue,          
    citecolor=green,        
    filecolor=blue,      
    urlcolor=blue           
}
\usepackage{hhline}
\usepackage{multirow}
\usepackage{makecell}
\usepackage{stfloats}
\usepackage{textcomp}
\usepackage{url}
\usepackage{units} 
\usepackage{verbatim}

\usepackage[caption=false, font=footnotesize]{subfig}
\usepackage[printonlyused]{acronym}
\usepackage[table]{xcolor}

\usepackage{algorithm}
\usepackage{algpseudocode}
\usepackage{tikz} 
\usepackage[utf8]{inputenc}
\usepackage{pgfplots} 
\pgfplotsset{width=10cm,compat=1.9}
\usepackage{pgfplotstable}
\usepackage{xurl} 
\usepackage[colorinlistoftodos,bordercolor=orange,backgroundcolor=orange!20,linecolor=orange,textsize=scriptsize]{todonotes}
\usepackage{yfonts}

\algtext*{EndWhile}
\algtext*{EndIf}
\algtext*{EndFor}
\usetikzlibrary {arrows.meta}
\makeatletter

\newtheorem{lem}{Lemma}

\newtheorem{proper}{Property}

\makeatletter
\algnewcommand{\LineComment}[1]{\Statex \hskip\ALG@thistlm \(\triangleright\) #1}
\makeatother

\newacro{aoa} [AoA] {angle-of-arrival}
\newacro{aod} [AoD] {angle-of-departure}
\newacro{amp} [AMP] {approximate message passing}
\newacro{arv} [ARV] {array response vector}
\newacro{awgn} [AWGN] {additive white Gaussian noise}
\newacro{cdf} [CDF] {cumulative distribution function}
\newacro{cf} [CF] {cell-free}
\newacro{ckm} [CKM] {channel knowledge map}
\newacro{cp} [CP] {cyclic prefix}
\newacro{crb} [CRB] {Cram{\'e}r-Rao bound}
\newacro{crlb} [CRLB] {Cram{\'e}r-Rao lower bound}
\newacro{cs} [CS] {compressed sensing}
\newacro{csi} [CSI] {channel state information}
\newacro{dft}[DFT]{discrete Fourier transform}
\newacro{em} [EM] {expectation-maximization}
\newacro{eer} [EER] {equal error rate}
\newacro{fa} [FA] {false alarm}
\newacro{fim} [FIM] {Fisher information matrix}
\newacro{ghz} [GHz] {gigahertz}
\newacro{glrt} [GLRT]{generalized likelihood ratio test}
\newacro{iid}[i.i.d.]{independently and identically distributed}
\newacro{isac} [ISAC] {integrated sensing and communications}
\newacro{kld} [KLD] {Kullback–Leibler divergence}
\newacro{llr}[LLR]{log-likelihood ratio}
\newacro{lmmse} [LMMSE] {linear minimum mean-square error}
\newacro{los} [LoS] {line-of-sight}
\newacro{ls} [LS] {least-square}
\newacro{lse} [LSE] {least-square estimator}
\newacro{lsfc} [LSFC] {large scale fading coefficient}
\newacro{lb} [LB] {lower bound}
\newacro{map} [MAP] {maximum a posteriori}
\newacro{md} [MD] {misdetection}
\newacro{mf} [MF] {matched filter}
\newacro{ml} [ML] {maximum likelihood}
\newacro{mle} [MLE] {maximum likelihood estimation}
\newacro{mmle} [MMLE] {mismatched maximum likelihood estimation}
\newacro{mcrb} [MCRB] {misspecified Cram\'er-Rao bound}
\newacro{mcrlb} [MCRLB] {misspecified Cram\'er-Rao lower bound}
\newacro{mse} [MSE] {mean-square error}
\newacro{mimo} [MIMO] {multiple-input multiple-output}
\newacro{mmwave} [mmWave] {millimeter-wave}
\newacro{mmse} [MMSE] {minimum mean-square error}
\newacro{nlos} [NLoS] {non-line-of-sight}
\newacro{nmse} [NMSE] {normalized mean squared error}
\newacro{np} [NP] {Neyman-Pearson}
\newacro{ofdm} [OFDM] {orthogonal frequency-division multiplexing}
\newacro{pdf}[PDF]{probability density function}
\newacro{rv} [RV] {random variable}
\newacro{rach} [RACH] {random access channel}
\newacro{rm} [RM]{radio map}
\newacro{rss} [RSS] {received signal strength}
\newacro{ru} [RU] {radio unit}
\newacro{ris} [RIS] {reconfigurable intelligent surface}
\newacro{simo} [SIMO] {single-input multiple-output}
\newacro{se} [SE] {state evolution}
\newacro{svd} [SVD] {singular value decomposition}
\newacro{snr} [SNR] {signal-to-noise ratio}
\newacro{tdoa} [TDoA] {time-difference-of-arrival}
\newacro{toa} [ToA] {time-of-arrival}
\newacro{ue} [UE] {user equipment}
\newacro{ula} [ULA] {uniform linear array}
\newacro{upa} [UPA] {uniform planar array}
\newacro{ura} [uRA] {unsourced random access}
\newacro{zc} [ZC] {Zadoff--Chu}

\newcommand{\herm}{{\sf H}}

\newcommand{\transp}{{\sf T}}
\newcommand{\Tran}{{\sf T}}

\newcommand{\SNR}{{\sf SNR}}

\newcommand{\abs}[1]{\left|{#1}\right|}

\newcommand{\diag}{{\hbox{diag}}}
\newcommand{\blkdiag}{{\hbox{blkdiag}}}

\newcommand{\eqdef}{\stackrel{\Delta}{=}}

\newcommand{\nth}[1]{{#1}{\text{th}}}

\newcommand{\trace}{{\hbox{tr}}}

\DeclareMathOperator{\Circ}{Circ}

\renewcommand{\vec}{{\rm vec}}

\newcommand{\cp}{\mathrm{cp}}

\newcommand{\fd}{\mathrm{fd}}
\newcommand{\iid}{\text{i.i.d.}}
\newcommand{\los}{\mathrm{L}}

\newcommand{\mf}{\mathrm{MF}}

\newcommand{\nlos}{\mathrm{N}}

\newcommand{\se}{\mathrm{SE}}
\newcommand{\samp}{\mathrm{s}}
\newcommand{\td}{\mathrm{td}}

\newcommand{\thr}{\mathrm{th}}

\newcommand{\refsub}{\mathrm{ref}}

\newcommand{\av}{{\bf a}}
\newcommand{\bv}{{\bf b}}

\newcommand{\ev}{{\bf e}}

\newcommand{\gv}{{\bf g}}
\newcommand{\hv}{{\bf h}}

\newcommand{\mv}{{\bf m}}

\newcommand{\pv}{{\bf p}}

\newcommand{\rv}{{\bf r}}
\newcommand{\sv}{{\bf s}}

\newcommand{\uv}{{\bf u}}
\newcommand{\wv}{{\bf w}}
\newcommand{\vv}{{\bf v}}
\newcommand{\xv}{{\bf x}}
\newcommand{\yv}{{\bf y}}
\newcommand{\zv}{{\bf z}}
\newcommand{\zerov}{{\bf 0}}

\newcommand{\Am}{{\bf A}}
\newcommand{\Bm}{{\bf B}}
\newcommand{\Cm}{{\bf C}}
\newcommand{\Dm}{{\bf D}}

\newcommand{\Fm}{{\bf F}}

\newcommand{\Hm}{{\bf H}}
\newcommand{\Id}{{\bf I}}
\newcommand{\Jm}{{\bf J}}
\newcommand{\Km}{{\bf K}}

\newcommand{\Rm}{{\bf R}}
\newcommand{\Sm}{{\bf S}}

\newcommand{\Wm}{{\bf W}}

\newcommand{\Xm}{{\bf X}}
\newcommand{\Ym}{{\bf Y}}

\newcommand{\Rrm}{{\rm R}}
\newcommand{\Srm}{{\rm S}}

\newcommand{\Urm}{{\rm U}}

\newcommand{\Acal}{{\cal A}}

\newcommand{\Cc}{{\cal C}}
\newcommand{\Dc}{{\cal D}}
\newcommand{\Ec}{{\cal E}}

\newcommand{\Gc}{{\cal G}}
\newcommand{\Hc}{{\cal H}}

\newcommand{\Lc}{{\cal L}}

\newcommand{\Nc}{{\cal N}}

\newcommand{\Pc}{{\cal P}}

\newcommand{\Rc}{{\cal R}}

\newcommand{\Uc}{{\cal U}}

\newcommand{\Vc}{{\cal V}}

\newcommand{\muv}{\hbox{\boldmath$\mu$}}

\newcommand{\psiv}{\hbox{\boldmath$\psi$}}

\newcommand{\Gammam}{\hbox{\boldmath$\Gamma$}}

\newcommand{\Sigmam}{\hbox{\boldmath$\Sigma$}}
\newcommand{\Phim}{\hbox{\boldmath$\Phi$}}

\newcommand{\Psim}{\hbox{\boldmath$\Psi$}}

\newfont{\bb}{msbm10 scaled 1100}
\newcommand{\CC}{\mbox{\bb C}}
\newcommand{\PP}{\mbox{\bb P}}
\newcommand{\RR}{\mbox{\bb R}}

\newcommand{\EE}{\mbox{\bb E}}

\newcommand{\II}{\mbox{\bb I}}

\newcommand{\onebb}{\mathbbm{1}}

\newcommand{\csf}{{\sf c}}

\newcommand{\xsf}{{\sf x}}
\newcommand{\ysf}{{\sf y}}
\newcommand{\zsf}{{\sf z}}